\RequirePackage{fix-cm}
\documentclass[a4,twocolumn,epjc3,comma,sort&compress,natbib]{svjour3}
\bibpunct{[}{]}{,}{n}{}{,} 

\journalname{Eur. Phys. J. C}

\usepackage{latexsym}
\usepackage{amsmath}
\usepackage{amssymb}
\usepackage{amsfonts}

\usepackage[mathscr,scaled=1.15]{urwchancal}
\DeclareFontFamily{OT1}{pzc}{}
\DeclareFontShape{OT1}{pzc}{m}{it}%
{<-> s * [1.15] pzcmi7t}{}
\DeclareMathAlphabet{\mathpzc}{OT1}{pzc}{m}{it}

\usepackage{color}

\usepackage{supertabular}
\usepackage{placeins}
\usepackage{epsfig}
\usepackage{graphicx}

\definecolor{purple}{rgb}{0.5,0,0.5}
\definecolor{blue}{rgb}{0.0,0,0.9}
\definecolor{prdblue}{rgb}{0.133,0.118,0.498}
\usepackage[colorlinks=true, pdfstartview=FitV, linkcolor=prdblue, citecolor= prdblue, urlcolor=prdblue]{hyperref}

\begin{document}


\title{$\,$\\[-7ex]\hspace*{\fill}{\normalsize{\sf\emph{Preprint no}. NJU-INP 025/20}}\\[1ex]
Contact interaction analysis of pion GTMDs}

\author{Jin-Li Zhang\thanksref{eJLZ,NJNU}
        \and
       Zhu-Fang Cui\thanksref{eZFC,NJU,INP} 
       \and
       Jia-Lun Ping\thanksref{eJLP,NJNU} 
       \and
       Craig D.~Roberts\thanksref{eCDR,NJU,INP}
}

\thankstext{eJLZ}{jlzhang@njnu.edu.cn}
\thankstext{eZFC}{phycui@nju.edu.cn}
\thankstext{eJLP}{jlping@njnu.edu.cn}
\thankstext{eCDR}{cdroberts@nju.edu.cn}

\authorrunning{Jin-Li Zhang \emph{et al}.} 

\institute{Department of Physics, Nanjing Normal University, Nanjing 210023, China \label{NJNU}
            \and
            School of Physics, Nanjing University, Nanjing, Jiangsu 210093, China \label{NJU}
           \and
           Institute for Nonperturbative Physics, Nanjing University, Nanjing, Jiangsu 210093, China \label{INP}
            }

\date{23 September 2020}

\maketitle

\begin{abstract}
A contact interaction is used to calculate an array of pion twist-two, -three and -four generalised transverse light-front momentum dependent parton distribution functions (GTMDs).
Despite the interaction's simplicity, many of the results are physically relevant, amongst them a statement that GTMD size and shape are largely prescribed by the scale of emergent hadronic mass.
Moreover, proceeding from GTMDs to generalised parton distributions (GPDs), it is found that the pion's mass distribution form factor is harder than its electromagnetic form factor, which is harder than the gravitational pressure distribution form factor; the pressure in the neighbourhood of the pion's core is commensurate with that at the centre of a neutron star; the shear pressure is maximal when confinement forces become dominant within the pion; and the spatial distribution of transversely polarised quarks within the pion is asymmetric.
Regarding transverse momentum dependent distribution functions (TMDs), their magnitude and domain of support decrease with increasing twist.
The simplest Wigner distribution associated with the pion's twist-two dressed-quark GTMD is sharply peaked on the kinematic domain associated with valence-quark dominance; has a domain of negative support; and broadens as the transverse position variable increases in magnitude.
\end{abstract}


\section{Introduction}
\label{Introduction}
It is anticipated that an electron ion collider will be operating in the USA by 2030 \cite{Accardi:2012qut, Aguilar:2019teb}; construction of a similar machine is being discussed in China \cite{EicCWP, Chen:2020ijn}; new capabilities are expected at Conseil Europ{\'e}en pour la Recherche Nucl{\'e}aire (CERN) \cite{Denisov:2018unj}; and the Jefferson Laboratory (JLab) is currently operating at 12\,GeV \cite{McKeown:2019aun}.  Each of these facilities has given high priority to experiments that can yield data that may be used to draw three-dimensional (3D) images of hadrons, \emph{i.e}.\ measurements interpretable in terms of generalised or transverse momentum dependent parton distributions: GPDs or TMDs, respectively.

Hadron physics has long focused on one dimensional (1D) imaging of hadrons.  It is an ongoing effort, which remains crucial because many puzzles and controversies are unresolved.  For instance, even considering what may seem to be the simplest strong interaction systems \cite{Sufian:2019bol, Ding:2019qlr, Ding:2019lwe}, the valence-quark distribution in the pion attracts vigorous debate; the pion's glue and sea distributions are empirically unknown, with theoretical predictions only now becoming available; and kaon distributions are just beginning to receive renewed attention \cite{Lin:2020ssv, Cui:2020dlm, Cui:2020piK}.  The challenge of producing solid predictions for parton distributions within baryons is even greater.

Notwithstanding the need for new, precise data on 1D distributions and associated predictions with a traceable connection to quantum chromodynamics (QCD), the allure of GPDs and TMDs is difficult to resist, given that 3D imaging may enable entirely new aspects of hadron structure to be revealed.  Such functions serve as tools with which to probe the multidimensional structure of hadron light-front wave functions (LFWF), thereby providing access to, \emph{inter alia}: the distributions of mass, pressure and spin within a hadron, both in longitudinal and transverse directions; the sharing of these qualities amongst the various bound-state constituents; and to the spacetime volumes occupied by these constituents, \emph{i.e}.\ to their potentially different  ``confinement'' radii.

It should be understood, however, that in order to fully capitalise on 3D imaging data obtained at modern and anticipated facilities, using it to understand the many correlated phenomena which emerge from strong interactions in QCD, methods must be developed that enable GPDs and TMDs to be calculated within frameworks that are mathematically linked to the fundamental theory.  To see the importance of this, one need look no further than the thirty year controversy over the pion's valence quark distribution function \cite{Sufian:2019bol, Ding:2019qlr, Ding:2019lwe, Cui:2020dlm, Cui:2020piK, Holt:2010vj, Aicher:2010cb, Barry:2018ort, Lan:2019rba, Novikov:2020snp, Chang:2020kjj}.

Herein we explore and illustrate the capacity of generalised parton correlation functions (GPCFs) \cite{Meissner:2008ay} to serve as a framework for the unified calculation of GPDs and TMDs.  As this is a first step, we choose to study the pion and work with a confining, symmetry-preserving treatment of a vector$\,\times\,$vector contact interaction (CI) as the foundation for our analysis \cite{GutierrezGuerrero:2010md}.  A merit of this approach is that, by enabling a largely algebraic treatment of relevant processes and quantities, it provides for an insightful assessment of all results.  Moreover, when considered judiciously \cite{GutierrezGuerrero:2010md, Roberts:2010rn, Roberts:2011wy, Chen:2012txa, Segovia:2013rca, Serna:2017nlr, Wang:2018kto, Yin:2019bxe}, such results may often be interpreted from a QCD perspective because this treatment of the CI preserves many qualities of the leading-order truncation of QCD's Dyson-Schwinger equations (DSEs), itself a sound approach to many hadron observables \cite{Horn:2016rip, Eichmann:2016yit, Burkert:2017djo, Fischer:2018sdj, Roberts:2020hiw, Qin:2020rad, Barabanov:2020jvn}.

Our analysis begins in Sec.\,\ref{SecGPCF} with a brief review of the GPCF for a $J=0$ hadron.
Section~\ref{SecCI}, augmented by \ref{AppFormulae}, then describes our CI treatment of the pion and its coupling to photons.
The pion GPCF is used in Sec.\,\ref{SecT2GTMDs} as the basis for calculating the four twist-two generalised transverse momentum dependent parton distribution functions (GTMDs) associated with dressed-quarks within the CI pion.  The discussion highlights the role played by emergent hadronic mass (EHM) in determining the properties of each GTMD.
(CI results for all twist-three and twist-four GTMDs are provided in \ref{AppendixT3} and \ref{AppendixT4}, respectively.)
In Sec.\,\ref{SecT2GPDs}, the twist-two GTMDs are integrated over their light-front-transverse momentum argument, $k_\perp^2$, to yield results for the pion's vector and tensor GPDs.  Features of the derived electromagnetic, gravitational, and transverse-spin distributions are also canvassed.
Section~\ref{SecTMDs} shows how one proceeds from GTMDs to TMDs.  It provides explicit formulae for all four TMDs supported by the CI in the absence of a model for the Wilson line and highlights their relative sizes and domains of $k_\perp^2$-support.
Section~\ref{SecWigner} highlights and illustrates the connection between GPCFs and Wigner distributions by presenting the CI result for a Wigner distribution associated with pion twist-two GPDs and TMDs.
A summary and perspective is provided in Sec.\,\ref{epilogue}.

\section{Generalised Parton Correlation Function}
\label{SecGPCF}
\begin{figure}[t]
\centerline{%
\includegraphics[clip, width=0.45\textwidth]{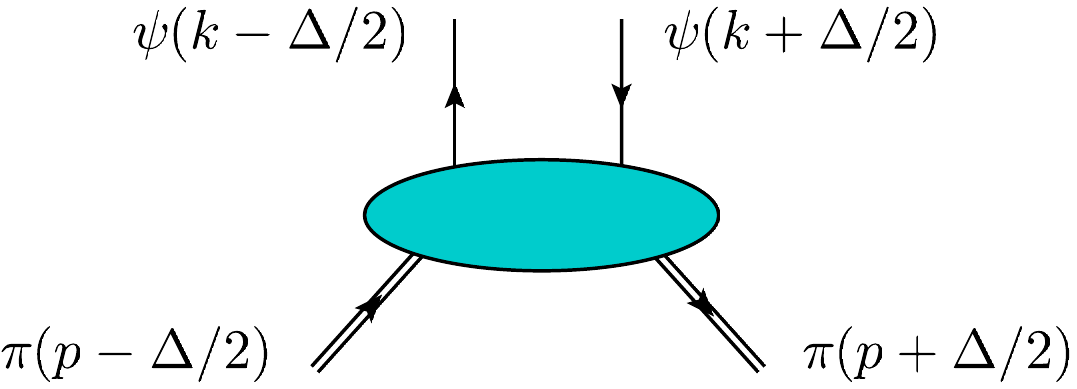}}
\caption{\label{GPCFkinematics}
Momentum-space conventions used in defining the in-pion quark-quark correlator in Eq.\,\eqref{piGPCF}.
}
\end{figure}

We begin by considering the following in-pion quark-quark correlator \cite{Meissner:2008ay}:
\begin{align}
&W_{ij}(P,k,\Delta,\bar N;\eta)  = \int\frac{d^4 z}{(2\pi)^4} \, {\rm e}^{i k\cdot z} \nonumber \\
& \times \langle\pi(p^\prime)|\,\bar\psi_j(-\tfrac{1}{2} z){\mathpzc W}(-\tfrac{1}{2} z,\tfrac{1}{2} z;\bar n)\,
\psi_i(\tfrac{1}{2} z) \,
|\pi(p)\rangle\,,
\label{piGPCF}
\end{align}
where:
\begin{equation}
P=(p^\prime+p)/2\,,\;
\Delta = p^\prime - p\,,\;
P\cdot \Delta = 0\,;
\end{equation}
and $k$ is the relative quark-antiquark momentum.  These conventions are illustrated in Fig.\,\ref{GPCFkinematics}.

The hitherto undefined quantity in Eq.\,\eqref{piGPCF} is the Wilson line, ${\mathpzc W}(-\tfrac{1}{2} z,\tfrac{1}{2} z;\bar n)$, where $\bar n$ is a light-like four-vector, $\bar n^2=0$, antiparallel to $P$, $\bar n\cdot P = P^-$, and the path is chosen as a sequence of line segments \cite{Meissner:2008ay, Collins:2002kn}:
\begin{equation}
-\frac{z}{2} \to -\frac{z}{2} + \frac{1}{\epsilon} \bar n \to \frac{z}{2} + \frac{1}{\epsilon} \bar n \to \frac{z}{2}\,,\quad \epsilon \to 0^+.
\end{equation}
The same path is achieved by rescaling $\bar n\to \lambda \bar n$, $\lambda \in \mathbb{R} $, $\lambda >0$; hence, with $\hat P^2=1$, Eq.\,\eqref{piGPCF} only depends on
\begin{equation}
\label{EqN}
\bar N = \bar n/\bar n\cdot \hat P\,.
\end{equation}
The quantity $\eta$ in Eq.\,\eqref{piGPCF} expresses the one remaining degree of freedom, \emph{viz}.\ $\eta = {\rm sign}(\bar n_0)$, in which case $\eta=\pm 1$ describe, respectively, future and past Wilson line trajectories.

One passes to generalised transverse-momentum dependent parton distribution functions (GTMDs) by first considering the following partially integrated quantity:
\begin{align}
&W_{ij}(P,x,\vec{k}_\perp,\Delta,N;\eta)=  \int\frac{d^4 z}{(2\pi)^4} \, {\rm e}^{i k\cdot z}\,\delta(n\cdot z) \nonumber \\
& \times  \langle\pi(p^\prime)|\,\bar\psi_j(-\tfrac{1}{2} z){\mathpzc W}(-\tfrac{1}{2} z,\tfrac{1}{2} z;\bar n)\,
\psi_i(\tfrac{1}{2} z) \,
|\pi(p)\rangle\,,
\label{piGPCF1}
\end{align}
where $n$ is a light-like four-vector for which $n\cdot P = P^+$.

The object in Eq.\,\eqref{piGPCF1} is a Dirac-matrix valued function and, as usual, contributions at various orders in a twist expansion can be obtained by appropriate projection operations.  Namely, with $\mathpzc H$ being some suitably chosen combination of Dirac matrices, then the scalar functions of interest -- the GTMDs -- are obtained via
\begin{align}
& W^{[{\mathpzc H}]}(P,x,\vec{k}_\perp,\Delta,N;\eta) = \frac{1}{2} W_{ij}(P,x,\vec{k}_\perp,\Delta,N;\eta) {\mathpzc H}_{ji}\nonumber \\
& =  \int\frac{d^4 z}{2(2\pi)^4} \, {\rm e}^{i k\cdot x}\,\delta(n\cdot z)  \nonumber \\
& \times \langle\pi(p^\prime)|\,\bar\psi_j(-\tfrac{1}{2} z){\mathpzc H}_{ji}{\mathpzc W}(-\tfrac{1}{2} z,\tfrac{1}{2} z;\bar n)\,
\psi_i(\tfrac{1}{2} z) \,
|\pi(p)\rangle\,. \label{GTMD}
\end{align}
Referring to Fig.\,\ref{GPCFkinematics}, this operation corresponds to the insertion of ${\mathpzc H}$ as a connection between the open quark and antiquark lines: $\psi(k\mp \Delta/2)$, respectively.

As defined by Eq.\,\eqref{GTMD}, a given GTMD is a complex-valued function: the real part is even under the time-reversal operation ($T$-even), whereas the imaginary part is $T$-odd.  Equally, they are even (odd) under $\eta \to -\eta$.  (Recall $\eta=\pm 1$ specifies the time-direction of the Wilson line used to define the GTMD.)  Beginning with Eq.\,\eqref{GTMD}, GPDs are obtained by integration over $\vec{k}_\perp$: only the $T$-even piece survives, which is independent of $\eta$; and the array of TMDs is obtained by setting $\Delta =0$, which entails $\xi=0$.

\section{Contact Interaction}
\label{SecCI}
When formulating the continuum bound-state problem for hadrons, the basic element is the quark-quark scattering kernel; and at leading-order in the symmetry-preserving truncation scheme introduced in Refs.\,\cite{Munczek:1994zz, Bender:1996bb}, that is the rainbow-ladder (RL) kernel ($k = p_1-p_1^\prime = p_2^\prime -p_2$):\footnote{\label{fn1}From this point onwards, we use the Euclidean metric and Dirac-matrix conventions described in Ref.\,\cite[Appendix~A]{Chen:2012qr}.}
\begin{subequations}
\label{KDinteraction}
\begin{align}
\mathscr{K}_{\alpha_1\alpha_1',\alpha_2\alpha_2'}  & = {\mathpzc G}_{\mu\nu}(k) [i\gamma_\mu]_{\alpha_1\alpha_1'} [i\gamma_\nu]_{\alpha_2\alpha_2'}\,,\\
 {\mathpzc G}_{\mu\nu}(k)  & = \tilde{\mathpzc G}(k^2) T_{\mu\nu}(k)\,,
\end{align}
\end{subequations}
where $k^2 T_{\mu\nu}(k) = k^2 \delta_{\mu\nu} - k_\mu k_\nu$.
The key element is $\tilde{\mathpzc G}$; and two decades of study, using a combination of continuum and lattice methods \cite{Binosi:2014aea, Binosi:2016nme, Rodriguez-Quintero:2018wma, Cui:2019dwv}, have revealed that owing to the dynamical generation of a gluon mass-scale in QCD \cite{Cornwall:1981zr, Dudal:2003by, Bowman:2004jm, Aguilar:2008xm, RodriguezQuintero:2010wy, Boucaud:2011ug, Strauss:2012dg, Binosi:2014aea, Aguilar:2015bud, Siringo:2015wtx, Cyrol:2016tym, Gao:2017uox, Binosi:2019ecz}, $\tilde{\mathpzc G}$ saturates at infrared momenta:
\begin{align}
\tilde{\mathpzc G}(k^2) & \stackrel{k^2 \simeq 0}{=} \frac{4\pi \alpha_{0}}{m_G^2}\,.
\end{align}
In QCD \cite{Cui:2019dwv}: $\alpha_{\rm 0} \approx \pi$ and $m_G \approx 0.5\,{\rm GeV} \approx m_N/2$, where $m_N$ is the nucleon mass.

To proceed, we follow Ref.\,\cite{Yin:2019bxe}; namely, retaining $m_G=0.5\,$GeV but setting $\alpha_0/\pi = 0.36$.  This combination ensures a good description of $\pi$-meson properties.  Furthermore, since a momentum-independent interaction cannot support relative momentum between bound-state constituents, we simplify the tensor structure in Eqs.\,\eqref{KDinteraction}, defining the CI RL kernel as follows:
\begin{align}
\label{KCI}
\mathscr{K}_{\alpha_1\alpha_1',\alpha_2\alpha_2'}^{\rm CI}  & = \frac{4\pi \alpha_{0}}{m_G^2}
 [i\gamma_\mu]_{\alpha_1\alpha_1'} [i\gamma_\mu]_{\alpha_2\alpha_2'}\,.
 \end{align}

When using Eq.\,\eqref{KCI} in a DSE, it is necessary to impose an ultraviolet regularisation scheme.  It should be symmetry preserving so that the results maintain a meaningful connection with the Standard Model.  Moreover, since a CI does not produce a renormalisable theory, the associated regularisation mass-scale, $\Lambda_{\rm uv}$, is an additional physical parameter.  It may be interpreted as an upper bound on the momentum domain within which the properties of the associated system are practically momentum-independent.

As the final step in defining the CI, we include an infrared regularisation scale, $\Lambda_{\rm ir}$, when computing all integrals connected with bound-state problems \cite{Ebert:1996vx}.  Since chiral symmetry is dynamically broken by Eq.\,\eqref{KCI}, ensuring the absence of infrared divergences, $\Lambda_{\rm ir}$ is not a necessary part of the CI's definition.  Notwithstanding that, by excising momenta $k<\Lambda_{\rm ir}$, one achieves a rudimentary expression of confinement via elimination of quark production thresholds \cite{Roberts:2007ji, Qin:2013ufa, Gao:2015kea, Papavassiliou:2015aga, Lowdon:2015fig, Lucha:2016vte, Binosi:2016xxu, Binosi:2019ecz}. A natural choice for this scale is $\Lambda_{\rm ir} \sim \Lambda_{\rm QCD}$.  We set $\Lambda_{\rm ir} = 0.24\,$GeV.

Assuming isospin symmetry, it only remains to fix the current-mass, $m$, of the light quarks.  That may be achieved by solving the pion bound state problem specified by the kernel in Eq.\,\eqref{KCI}.  In this case, the gap equation for the dressed light-quark propagator is
\begin{align}
\label{GapEqn}
S_f^{-1}(p) & = i\gamma\cdot p +m
+ \frac{16 \pi}{3} \frac{\alpha_{0}}{m_G^2}
\int \frac{d^4q}{(2\pi)^4} \gamma_\mu S(q) \gamma_\mu\,.
\end{align}
The integral is quadratically divergent.  When it is regularised in a Poincar\'e-invariant manner, the gap equation solution is
\begin{equation}
\label{genS}
S(p)^{-1} = i \gamma\cdot p + M\,,
\end{equation}
where $M$ is the dressed-quark mass, momentum-independent in the CI, determined by
\begin{equation}
M = m + M\frac{4\alpha_{0}}{3\pi m_G^2} \left[\int_0^\infty \!ds \, s\, \frac{1}{s+M^2}\right]_{\rm reg}\,.
\end{equation}

We define the regularised integral by writing \cite{Ebert:1996vx}
\begin{subequations}
\begin{align}
\nonumber
\frac{1}{s+M^2} & = \int_0^\infty d\tau\,{\rm e}^{-\tau (s+M^2)}  \\
&\rightarrow  \int_{\tau_{\rm uv}^2}^{\tau_{\rm ir}^2} d\tau\,{\rm e}^{-\tau (s+M^2)}
\label{RegC}\\
 &=
\frac{{\rm e}^{- (s+M^2)\tau_{\rm uv}^2}-{\rm e}^{-(s+M^2) \tau_{\rm ir}^2}}{s+M^2} \,, \label{ExplicitRS}
\end{align}
\end{subequations}
where $\tau_{\rm ir,uv}=1/\Lambda_{{\rm ir},{\rm uv}}$ are, respectively, the infrared and ultraviolet regulators described above.  Consequently, the gap equation becomes
\begin{equation}
M= m + M\frac{4\alpha_{0}}{3\pi m_G^2}\,\,{\cal C}_0(M^2)\,,
\label{gapactual}
\end{equation}
where
\begin{align}
\nonumber
{\cal C}_0(\sigma) &=
\int_0^\infty\! ds \, s \int_{\tau_{\rm uv}^2}^{\tau_{\rm ir}^2} d\tau\,{\rm e}^{-\tau (s+\sigma)}\\
& =
\sigma \big[\Gamma(-1,\sigma \tau_{\rm uv}^2) - \Gamma(-1,\sigma \tau_{\rm ir}^2)\big],
\label{eq:C0}
\end{align}
with $\Gamma(\alpha,y)$ being the incomplete gamma-function.

In an internally consistent treatment of a vector$\,\times\,$vector CI, the Bethe-Salpeter amplitude for the $\pi$-meson has the following form \cite{GutierrezGuerrero:2010md, Roberts:2011wy, Chen:2012txa}:
\begin{align}
\label{BSAcontactpion}
\Gamma_{\pi}(Q) = \gamma_5 \left[ i E_{\pi}(Q) + \frac{1}{M}\gamma\cdot Q F_{\pi}(Q)\right]\,.
\end{align}
Here, $Q$ is the pion's total momentum, $Q^2 = -m_{\pi}^2$, $m_{\pi}$ is the pion mass; $M$ is obtained from the contact-interaction gap equation, Eq.\,\eqref{gapactual}; and $E_\pi$, $F_\pi$ do not depend on the relative quark-antiquark momentum.

The amplitude, $\Gamma_\pi$, is obtained from the following homogeneous Bethe-Salpeter equation:
\begin{equation}
\Gamma_{\pi}(Q) =  - \frac{16 \pi}{3} \frac{\alpha_{0}}{m_G^2}
\int \! \frac{d^4 \ell}{(2\pi)^4} \gamma_\mu S(\ell+Q) \Gamma_{\pi}(Q)S(\ell) \gamma_\mu \,.
\label{LBSEI}
\end{equation}
Employing the symmetry-preserving regularisation scheme of Refs.\,\cite{GutierrezGuerrero:2010md, Chen:2012txa}, which emulates dimensional regularisation and requires
\begin{equation}
0 = \int_0^1d\alpha \,
\big[ {\cal C}_0(\omega(\alpha,Q^2))
%
+ \, {\cal C}_1(\omega(\alpha,Q^2))\big], \label{avwtiP}
\end{equation}
where ${\cal C}_1$ is given in Eqs.\,\eqref{Cndef}, \eqref{C012def} and
\begin{align}
\omega(\alpha,Q^2) &= M^2 + \alpha \bar\alpha Q^2\,,
\label{eq:omega}
\end{align}
$\bar \alpha = 1-\alpha$, one arrives at the following pair of coupled equations:
\begin{equation}
\label{bsefinalE}
\left[
\begin{array}{c}
E_{\pi}(Q)\\
F_{\pi}(Q)
\end{array}
\right]
= \frac{4 \alpha_{0}}{3\pi m_G^2}
\left[
\begin{array}{cc}
{\cal K}_{EE}^{\pi} & {\cal K}_{EF}^{\pi} \\
{\cal K}_{FE}^{\pi} & {\cal K}_{FF}^{\pi}
\end{array}\right]
\left[\begin{array}{c}
E_{\pi}(Q)\\
F_{\pi}(Q)
\end{array}
\right],
\end{equation}
with the matrix elements $\{ {\cal K}_{EE}^{\pi}, {\cal K}_{EF}^{\pi}, {\cal K}_{FE}^{\pi}, {\cal K}_{FF}^{\pi}\}$ defined in Eqs.\,\eqref{fgKernel}.  Evidently, the kernel is only defined after the gap equation has been solved.

Inspection of Eqs.\,\eqref{bsefinalE}, \eqref{fgKernel} reveals that a nonzero value for $E_\pi$ enforces $F_\pi \neq 0$, \emph{i.e}.\ any theory with a traceable connection to a vector-boson exchange interaction must retain both $E_\pi$, $F_\pi$.  (When the interaction is momentum dependent, then two other amplitudes are also nonzero \cite{LlewellynSmith:1969az, Maris:1997tm}.)  If $F_\pi$ is omitted, then one arrives at a model, which although it may be useful for parametrising data, cannot contribute to the development of insights into characteristics of the Standard Model's Nambu-Goldstone modes \cite{Roberts:2010rn, Chen:2012txa}.

Eq.\,\eqref{bsefinalE} is an eigenvalue problem.  It has a solution when $Q^2=-m_\pi^2$, at which point the eigenvector is the meson's Bethe-Salpeter amplitude.  Working with the on-shell solution, normalised canonically according to Eqs.\,\eqref{normcan}, \eqref{normcan2}, the pion's leptonic decay constant is given by ($N_c=3$):
\begin{equation}
f_{\pi} = \frac{N_c}{2\pi^2}\frac{1}{ M}\,
\big[ E_{\pi} {\cal K}_{FE}^{\pi} + F_{\pi}{\cal K}_{FF}^{\pi} \big]_{Q^2=-m_{\pi}^2}\,. \label{ffg}
\end{equation}
In the chiral limit, \emph{i.e}.\ using solutions obtained with $m=0$ in Eq.\,\eqref{GapEqn}, this reduces to \cite{GutierrezGuerrero:2010md}
\begin{equation}
f_{\pi}^0 = \frac{N_c}{4\pi^2}\frac{1}{ M}\,{\cal C}_1(M^2)  [E_\pi - 2 F_\pi]\,.
\label{ffg0}
\end{equation}

\begin{table}[t]
\caption{\label{pionresults}
With input parameters \cite{Roberts:2011wy, Yin:2019bxe} $m_G=0.5\,$GeV, $\alpha_0 = 0.36\pi$, $\Lambda_{\rm ir} = 0.24\,$GeV, $\Lambda_{\rm uv} = 0.905\,$GeV, solving the coupled gap and Bethe-Salpeter equations yields the results listed here.
(Dimensioned quantities in GeV.)}
\begin{center}
\begin{tabular*}
{\hsize}
{
c@{\extracolsep{0ptplus1fil}}
c@{\extracolsep{0ptplus1fil}}
c@{\extracolsep{0ptplus1fil}}
c@{\extracolsep{0ptplus1fil}}
c@{\extracolsep{0ptplus1fil}}
c@{\extracolsep{0ptplus1fil}}}\hline
$m$ & $M$ & $m_\pi$ & $f_\pi$ & $E_\pi$ & $F_\pi$ \\
$0.007\ $& $0.368\ $ & $0.14\ $ & $0.10\ $ & $3.64\ $ & $0.481\ $\\\hline
\end{tabular*}
\end{center}
\end{table}

Solving Eqs.\,\eqref{GapEqn}, \eqref{bsefinalE}, we obtain the results listed in Table~\ref{pionresults}, reproducing those reported elsewhere \cite{Roberts:2011wy, Yin:2019bxe}.

For subsequent use, here we also introduce the dressed photon-quark vertex, $\Gamma_\mu^\gamma$.  Using Eq.\,\eqref{KCI}, one has
\begin{align}
&\Gamma_\mu^\gamma(\Delta) = \gamma_\mu \nonumber \\
&- \frac{16\pi\alpha_0}{3m_G^2} \int\frac{d^4 \ell}{(2\pi)^4}
\gamma_\alpha \, S(\ell_{+\Delta}) \, \Gamma_\mu^\gamma(\Delta) \, S(\ell_{-\Delta}) \gamma_\alpha\,.
\end{align}
Owing to the vector Ward-Green-Takahashi identity (WGTI), preserved in our regularisation of the contact interaction, the solution takes the form \cite{Roberts:2010rn}
\begin{equation}
 \Gamma_\mu^\gamma(\Delta) = \gamma_\mu^{\rm T} P_{\rm T}(\Delta^2) + \gamma_\mu^{\rm L}\,,
\end{equation}
where $\Delta\cdot\gamma_\mu^{\rm T}=0$, $\gamma_\mu^{\rm T} +\gamma_\mu^{\rm L} = \gamma_\mu$,
\begin{subequations}
\begin{align}
P_{\rm T}(\Delta^2) & = \frac{1}{1+K_\gamma(\Delta^2)}\,, \label{EqPT}\\
K_\gamma(\Delta^2) & = \frac{4\alpha_0\Delta^2}{3\pi m_G^2}
\int_0^1 d\alpha\,\alpha\bar\alpha\,\bar{\cal C}_1(\omega(\alpha,\Delta^2)\,.
\end{align}
\end{subequations}
As expected of RL truncation studies of the photon-quark vertex \cite{Maris:1999bh, Roberts:2000aa}, the dressing function, $P_{\rm T}(\Delta^2)$, exhibits a simple pole at $\Delta^2=-m_\rho^2$, where $m_\rho$ is the mass of the $\rho$-meson that is generated by the interaction.

\section{Pion twist-two GTMDs}
\label{SecT2GTMDs}
There are three twist-two pion GTMDs.  They are obtained with the following choices in Eq.\,\eqref{GTMD}:
\begin{align}
{\mathpzc H} \to \{ {\mathpzc H}_1 = i n\cdot\gamma \,,\,
{\mathpzc H}_2 = i n\cdot \gamma \gamma_5\,,\,
{\mathpzc H}_3 = i \sigma_{j\mu} n_\mu \}.
\end{align}
The simplest is that associated with ${\mathpzc H}_1$, which relates to the pion valence-quark distribution function and electromagnetic form factor.  We therefore use it to illustrate the computational techniques.

Mapping into Euclidean metric:
\begin{equation}
W^{[{\mathpzc H}_1]}(P,x,\vec{k}_\perp,\Delta,N;\eta) \to F_1(x,k_\perp^2,\xi,t) \,;
\end{equation}
and since a RL truncation was used to solve the Bethe-Salpeter equation, then internal consistency and preservation of symmetries requires a kindred truncation for the GTMD, in which case
\begin{align}
& F_1(x,k_\perp^2,\xi,t)  = 2 N_c {\rm tr}_{\rm D}\int \frac{dk_3 dk_4}{(2\pi)^4} \delta_n^x(k) \,\Gamma_\pi(-p^\prime)\, \nonumber \\
& \times S(k_{+\Delta})\, n\cdot \Gamma_\gamma(\Delta)\, S(k_{-\Delta})\,\Gamma_\pi(p)\, S(k-P)\,,
\label{F1start}
\end{align}
where ${\rm tr}_{\rm D}$ indicates a trace over spinor indices,
$\delta_n^x(k) = \delta(n\cdot k - x n\cdot P)$,
\begin{equation}
k_{\pm \Delta} = k\pm \Delta/2\,\;
t=-\Delta^2,\;
p\cdot\Delta = -\Delta^2/2= - p^\prime\cdot\Delta\,,
\label{EqKinematics}
\end{equation}
and the ``skewness'' $\xi = [-n\cdot \Delta]/[2 n\cdot P]$, $|\xi|\leq 1$.

Two observations are important here.
(\emph{A}).\ When using a contact interaction, Eq.\,\eqref{KCI}, the pion Bethe-Salpeter amplitude is independent of relative momentum, Eq.\,\eqref{BSAcontactpion}.  Hence, on ${\mathpzc D} = \{ x\,|\,x < - \xi \, \cup \, x>\xi\ \, \cap |x|\leq 1\}$, the leading-twist corrections to Eq.\,\eqref{F1start} that were identified in Ref.\,\cite{Chang:2014lva} and exploited in Refs.\,\cite{Ding:2019qlr, Ding:2019lwe, Cui:2020dlm, Cui:2020piK} can be neglected.  However, additional contributions should be considered on the complementary domain, ${\mathpzc E} = \{ x\,|\, - \xi  < x < \xi \}$ \cite{Theussl:2002xp}.
(\emph{B}).\ Using a realistic, momentum-dependent interaction, the analogue of Eq.\,\eqref{F1start} can be a useful approximation to the pion GTMD at an hadronic scale, $\zeta_H<\Lambda_{\rm uv}$, at which the dressed quasiparticles obtained as solutions to the quark gap equation express all properties of the bound state under consideration, \emph{e.g}.\ they carry all the hadron's momentum at $\zeta_H$.  In this case \cite{Ding:2019qlr, Ding:2019lwe, Raya:2015gva, Horn:2016rip, Gao:2017mmp, Ding:2018xwy, Cui:2020dlm, Cui:2020piK}, predictions appropriate to experiments at $\zeta>\zeta_H$ are obtained using the $\zeta$-evolution equations appropriate to the distribution under consideration \cite{Dokshitzer:1977, Gribov:1972, Lipatov:1974qm, Altarelli:1977, Lepage:1979zb, Efremov:1979qk, Lepage:1980fj, Aybat:2011zv}.   Despite the fact that the contact interaction does not define a renormalisable model, we maintain this perspective herein.

In proceeding with a WGTI-preserving evaluation of Eq.\,\eqref{F1start}, we first compute the spinor trace; then using the following identities [$D(k^2)=k^2+M^2$]:
\begin{subequations}
\begin{align}
2 k\cdot p & = D(k_{-\Delta}^2)-D((k-P)^2)+P^2 - \Delta^2/4\,,\\
2 k\cdot p^\prime & = D(k_{+\Delta}^2)-D((k-P)^2)+P^2 - \Delta^2/4\,,\\
2 k^2 & = D(k_{+\Delta}^2)+D(k_{-\Delta}^2) - 2M^2 - \Delta^2/2\,,\\
2 k\cdot \Delta & = D(k_{+\Delta}^2) - D(k_{-\Delta}^2)]\,,
\end{align}
\end{subequations}
cancel each common numerator and denominator factor; and finally use Feynman parametrisations to simplify all remaining denominators.  In this way, one arrives at
\begin{align}
F_1(x,& k_\perp^2,\xi,t)  =  \nonumber \\
& \times \bar P_{\rm T} \left[E_\pi^2 \,F_1^{EE} +  E_\pi F_\pi \, F_1^{EF}  + F_\pi^2 \,F_1^{FF}\right] \,,
\label{EqGTMDF1}
\end{align}
where $\bar P_{\rm T}= [\theta_{\bar\xi\xi} + P_{\rm T}(-t)(1-\theta_{\bar\xi\xi}) ] $ and $(r=k_\perp^2)$:
\begin{subequations}
\begin{align}
 F_1^{EE}(x,r,\xi,t)  &= T_1^{EE} + T_2^{EE} \nonumber \\
&  \quad + \frac{N_c}{8\pi^3} \frac{1}{\sigma_2^r}\bar {\cal C}_2(\sigma_2^{r})\frac{\theta_{\bar\xi\xi} x}{\xi}\,, \\
 F_1^{EF}(x,r,\xi,t)  & = - 2 \, T_1^{EE} - 4 T_2^{EE} \,,\\
 F_1^{FF}(x,r,\xi,t)  & =4 T_2^{EE} \nonumber \\
 & \quad - \frac{N_c}{8\pi^3} \frac{1}{\sigma_2^r}\bar {\cal C}_2(\sigma_2^{r}) \frac{\theta_{\bar\xi\xi}t}{M^2\xi}  \left[1-\frac{x^2}{\xi^2} \right]\,,
\end{align}
\end{subequations}
with
\begin{subequations}
\begin{align}
  T_1^{EE}& (x,r,\xi,t) = \nonumber \\
 & \frac{N_c}{4\pi^3} \left[
 \frac{\theta_{\bar \xi 1}}{\sigma_1^{r,1}}\bar {\cal C}_2(\sigma_1^{r,1})
 +  \frac{\theta_{\xi 1}}{\sigma_1^{r,-1}}\bar {\cal C}_2(\sigma_1^{r,-1}) \right] \,, \\
  T_2^{EE}&(x,r,\xi,t) = \nonumber \\
 &   \frac{3 N_c}{8\pi^3} \left[\frac{2x}{\xi} m_\pi^2 +\frac{1-x}{\xi}t\right] \int_0^1 d\alpha\,
\frac{\theta_{\alpha\xi}}{[\sigma_3^r]^2}\bar {\cal C}_3(\sigma_3^{r}) \,,
\end{align}
\end{subequations}
and
{\allowdisplaybreaks
\begin{subequations}
\label{Eqthetas}
\begin{align}
\theta_{\bar \xi 1} & = x\in[-\xi,1]\,,\\
\theta_{\xi 1} & = x\in[\xi,1]\,,\\
\theta_{\bar\xi \xi} & = x\in [-\xi,\xi]\,,\\
\theta_{\alpha\xi} & = x\in [\alpha(1+\xi)-\xi, \alpha(1-\xi)+\xi] \cap x\in[-1,1]\,.
\end{align}
\end{subequations}}
\hspace*{-0.4\parindent}For later use, we note that one can write
$\theta_{\bar\xi \xi}/\xi = \Theta(1-x^2/\xi^2)$, where $\Theta(x)$ is the Heaviside function,
and $\theta_{\alpha\xi}/\xi $ $= \Theta((1-\alpha)^2 - (x-\alpha)^2/\xi^2)\Theta(1-x^2)$.  Under $\xi \to -\xi$: $\theta_{\bar \xi 1} \leftrightarrow \theta_{\xi 1}$; and $\theta_{\bar\xi \xi}/\xi$, $\theta_{\alpha\xi}/\xi$ are invariant.

Here it is worth recalling a Goldberger-Treiman relation that emerges in a WGTI-preserving treatment of the CI.  Namely, in the absence of a Higgs mechanism -- so that $m=0$ in the gap equation, Eq.\,\eqref{GapEqn}, and one is dealing with the chiral limit \cite{GutierrezGuerrero:2010md}:
\begin{equation}
\label{EqGTR}
E_\pi^0 = \frac{M^0}{f_\pi^0}\,,
\end{equation}
where the superscript ``$0$'' indicates evaluation in the chiral-limit.
Both $M^0$ and $f_\pi^0$ are order parameters for dynamical chiral symmetry breaking (DCSB) \cite{Roberts:2000aa}, which itself is an expression of EHM in the Standard Model \cite{Roberts:2020hiw}.  Moreover, Eq.\,\eqref{EqGTR} is practically unchanged at physical light-quark current masses.  (Similar statements also hold in QCD \cite{Maris:1997hd, Qin:2014vya}.)
Consequently, the strength of the pion's canonically normalised Bethe-Salpeter amplitude is a direct measure of EHM; hence, the CI formulae presented above and those to follow reveal that the size and shape of every one of the pion's GTMDs are largely determined by the character of EHM.

Consider ${\mathpzc H}_2 = i n\cdot \gamma \gamma_5$ and define $\varepsilon^\perp_{ij} = \varepsilon_{\alpha \beta i j} \bar n_\alpha n_\beta$, then
\begin{equation}
W^{[{\mathpzc H}_2]}(P,x,\vec{k}_\perp,\Delta,N;\eta) \to
i \varepsilon^\perp_{ij} k_i \Delta_j
\tilde G(x,k_\perp^2,\xi,t) \,,
\end{equation}
where $(r=k_\perp^2)$:
\begin{subequations}
\begin{align}
 \tilde G_1&(x,r,\xi,t)  =
\frac{N_c}{4\pi^3\xi} \bar P_{\rm T}  \left[\frac{F_\pi^2}{M^2} \frac{\theta_{\bar\xi \xi}}{\sigma_2^r}\bar {\cal C}_2(\sigma_2^{r})
-  {\mathpzc R}(x,r,\xi,t) \right] ,\\
%
%
%
{\mathpzc R}&(x,r,\xi,t)  = 3N_{EF} \int_0^1 d\alpha\,\theta_{\alpha\xi}
\frac{1}{[\sigma_3^r]^2}{\cal C}_3(\sigma_3^{r}) \,,
\end{align}
\end{subequations}
with $N_{EF}=(E_\pi^2 - 4 E_\pi F_\pi +4 F_\pi^2)$ and, for subsequent use,
$\tilde N_{EF}=F_\pi(E_\pi - 2 F_\pi)$,
$\bar N_{EF}=F_\pi(E_\pi - F_\pi)$.

Insertion of ${\mathpzc H}_3 = i \sigma_{j\mu} n_\mu$ into Eq.\,\eqref{GTMD} produces two terms:
\begin{align}
W^{[{\mathpzc H}_3]}&(P,x,\vec{k}_\perp,\Delta,N;\eta) \nonumber \\
& \to
k_j H_1^k + \frac{n\cdot P \Delta_j - n\cdot \Delta P_j}{n\cdot P}H_1^\Delta\,,
\end{align}
where $(r=k_\perp^2)$:
{\allowdisplaybreaks
\begin{subequations}
\begin{align}
H_1^\Delta &(x,k_\perp^2,\xi,t) =
\bar P_{\rm T} \frac{N_c}{4\pi^3} \left[
- \frac{F_\pi^2}{M}  \theta_{\bar\xi\xi}\frac{1}{\xi}
\frac{{\mathpzc C}_2(\sigma_2^r)}{\sigma_2^r}\right. \nonumber \\
& \left. \quad + 3 N_{EF}
\int_0^1 \! d\alpha\, \theta_{\alpha\xi}\,\frac{M}{\xi} \frac{\bar{\mathpzc C}_3(\sigma_3^r)}{[\sigma_3^r]^2}\right]
\,,
\label{EqH1Delta}\\
H_1^k &(x,k_\perp^2,\xi,t) = \frac{N_c}{2 \pi^3}
\left[ \frac{\tilde N_{EF}}{M} \left(
\theta_{\xi 1}\frac{\bar{\mathpzc C}_2(\sigma_1^{r,-1})}{\sigma_1^{r,-1}}
\right. \right. \nonumber \\
& \left. \quad - \theta_{\bar\xi 1} \frac{\bar{\mathpzc C}_2(\sigma_1^{r,1})}{\sigma_1^{r,1}}\right)
\left. + \frac{2 \bar N_{EF}}{M} \theta_{\bar\xi\xi}
\frac{\bar{\mathpzc C}_2(\sigma_2^{r})}{\sigma_2^{r}}\right]\,.
\label{EqH1k}
\end{align}
\end{subequations}}
\hspace*{-0.5\parindent}Evidently, $H_1^k (x,k_\perp^2,\xi,t)$ vanishes unless one uses the complete pion Bethe-Salpeter amplitude in Eq.\,\eqref{BSAcontactpion}, \emph{i.e}.\ $F_\pi \neq 0$.
In this connection it is worth recalling that inspection of Eqs.\,\eqref{bsefinalE}, \eqref{fgKernel} shows that a nonzero value for $E_\pi$ forces $F_\pi \neq 0$, \emph{i.e}.\ the strength of $F_\pi$ is also set by EHM.

\section{Pion twist-two GPDs}
\label{SecT2GPDs}
\subsection{Algebraic Results}
As noted in closing Sec.\,\ref{SecGPCF}, one proceeds from a GTMD to a GPD by integrating over $\vec{k}_\perp$; and focusing first on the leading twist GTMDs, one therefrom obtains two GPDs: \begin{subequations}
\begin{align}
{\mathsf H}_\pi(x,\xi,t) & = \int d^2\vec{k}_\perp F_1(x, k_\perp^2,\xi,t) \,, \label{EqGPDF}\\
{\mathsf E}_{\pi}^{\rm T}(x,\xi,t) & = \int d^2\vec{k}_\perp H_1^\Delta(x, k_\perp^2,\xi,t) \label{EqGPDH} \,,
\end{align}
\end{subequations}
where ${\mathsf H}_\pi$, ${\mathsf E}_\pi^{\rm T}$ may respectively be called the vector (no spin-flip) and tensor (spin-flip) GPDs.  The former is directly related to the pion's elastic electromagnetic form factor and gravitational form factors (mass and pressure/stress) \cite{Diehl:2003ny}, whereas the latter provides access to the dependence of the pion's quark distributions on their polarisation perpendicular to the pion's direction of motion (transversity) \cite{Polyakov:2002yz, Polyakov:2018zvc}.

Inserting Eq.\,\eqref{EqGTMDF1} into Eq.\,\eqref{EqGPDF} yields
\begin{equation}
{\mathsf H}_\pi (x,\xi,t)  = \bar P_{\rm T} \left[E_\pi^2 \,{\mathsf F}_1^{EE} +  E_\pi F_\pi \, {\mathsf F}_1^{EF}  + F_\pi^2 \,{\mathsf F}_1^{FF}\right] \,, \label{EqGPDFA}
\end{equation}
%
where
\begin{subequations}
\begin{align}
{\mathsf F}_1^{EE}(x,\xi,t) & = {\mathsf T}_1^{EE} +  {\mathsf T}_2^{EE}
+ \frac{N_c}{8\pi^2} \theta_{\bar\xi\xi} \frac{x}{\xi}\bar {\cal C}_1(\sigma_2^0)\,,\\
{\mathsf F}_1^{EF}(x,\xi,t) & =-2 {\mathsf T}_1^{EE} - 4 {\mathsf T}_2^{EE}\,,\\
{\mathsf F}_1^{FF}(x,\xi,t) & = 4 {\mathsf T}_2^{EE} \nonumber \\
& \quad -\frac{N_c}{16\pi^2}\bar {\cal C}_1(\sigma_2^{0}) \frac{\theta_{\bar\xi\xi} t }{\xi M^2 }
\left[ 1 - \frac{x^2}{\xi^2} \right],
\end{align}
\end{subequations}
with
\begin{subequations}
\label{DefintionsHpi}
\begin{align}
{\mathsf T}_1^{EE}(x,\xi) & = \frac{N_c}{8\pi^2}
\left[
\theta_{\bar\xi 1} \bar {\cal C}_1(\sigma_1^{0,1})
+ \theta_{\xi 1} \bar {\cal C}_1(\sigma_1^{0,-1})
\right], \\
{\mathsf T}_2^{EE}(x,\xi,t) & = \frac{N_c}{8\pi^2} [2x m_\pi^2+ (1-x)t ] \int_0^1 \! d\alpha\,
\frac{\theta_{\alpha\xi}}{\xi\sigma_3^0}\bar {\cal C}_2(\sigma_3^0)\,.
\end{align}
\end{subequations}

Using the results following Eq.\,\eqref{Eqthetas} and Eqs.\,\eqref{Eqsigmas}, it is straightforward to establish that
\begin{equation}
\label{TRinvariance}
{\mathsf H}_\pi(x,-\xi,t) = {\mathsf H}_\pi(x,\xi,t)\,,
\end{equation}
\emph{i.e}.\ our CI treatment preserves the time-reversal-invariance property of the GPD.

It is nevertheless deficient on the domain ${\mathpzc E} = \{ x\,|\, - \xi  < x < \xi \}$ because ${\mathsf H}_\pi(x,-\xi,t)$ does not satisfy the soft pion theorems \cite{Polyakov:1999gs} $(u=[1+x]/2)$:
\begin{subequations}
\label{EqSoft}
\begin{align}
{\mathsf H}_\pi(x,\xi,0) & = \tfrac{1}{2}\varphi_\pi(u) + {\rm O}(m_\pi^2)\,,\\
\int_{-1}^1dx\, {\mathsf H}_\pi(x,1,0) & =  {\rm O}(m_\pi^2)\,.
\end{align}
\end{subequations}
A remedy is described elsewhere \cite{Theussl:2002xp}; to wit, one must include interactions between the two pions in Fig.\,\ref{GPCFkinematics} that would lead to formation of a scalar meson-resonance.  Profiting from this understanding, we expand on the \emph{Ansatz} in Ref.\,\cite{Chouika:2017rzs} and correct the twist-two vector GPD:
\begin{subequations}
\label{HpiCorrected}
\begin{align}
{\mathsf H}_\pi(x,\xi,t) & \to \tilde {\mathsf H}_\pi(x,\xi,t) \\
& = {\mathsf H}_\pi(x,\xi,t) + {\mathsf D}_\pi(x,\xi,t) \,, \\
{\mathsf D}_\pi(x,\xi,t) & =[\tfrac{1}{2} {\mathsf H}_\pi(u,0,0)  - {\mathsf H}_\pi(x,\xi,0) ]
\xi^2 P_{\sigma}(t)\,, \label{EqDterm}
\end{align}
\end{subequations}
where $P_{\sigma}(t)$ is a quark+antiquark scalar-channel analogue of $P_{\rm T}(t)$ in Eq.\,\eqref{EqPT}.  It is readily established that $\tilde {\mathsf H}_\pi(x,\xi,t)$ is consistent with known mathematical GPD constraints and Eqs.\,\eqref{EqSoft}.\footnote{The factor $\xi^2$ in Eq.\,\eqref{EqDterm} should strictly be replaced by $\theta_{\bar\xi\xi}/p(\xi^2)$, where $p(\xi^2)$ is a simple polynomial, chosen to preserve GPD polynomiality; but that merely complicates numerical analysis without delivering practical improvement.}

Inserting Eq.\,\eqref{EqH1Delta} into Eq.\,\eqref{EqGPDH}, the twist-two tensor GPD is obtained:
\begin{align}
{\mathsf E}_\pi^{\rm T}(x,\xi,t)  & =
\bar P_{\rm T}(-t) \frac{N_c}{8\pi^2}
\left[
- \frac{F_\pi^2}{M}\frac{\theta_{\bar\xi\xi}}{\xi} \bar{\mathpzc C}_1(\sigma_2^0) \right.\nonumber \\
& \left. \quad + 2 M N_{EF} \int_0^1 d\alpha \frac{\theta_{\alpha\xi}}{\xi}
\frac{\bar{\mathpzc C}_2(\sigma_3^0)}{\sigma_3^0} \right]. \label{EqEpi}
\end{align}
The following remarks are pertinent:
$M {\mathsf E}_\pi^{\rm T}(x,\xi,t)$ is dimensionless;
relative to some other studies, \emph{e.g}.\ Refs.\,\cite{Brommel:2007xd, Fanelli:2016aqc}, our normalisation convention in Eq.\,\eqref{EqH1Delta} entails that ${\mathsf E}_\pi^{\rm T}(x,\xi,0)$ is nonzero in the chiral limit;
and once again using the results described in connection with Eqs.\,\eqref{Eqthetas}, \eqref{Eqsigmas}, one finds
\begin{equation}
{\mathsf E}_\pi^{\rm T}(x,-\xi,t) = {\mathsf E}_\pi^{\rm T}(x,\xi,t)\,.
\label{EqEpTRinvariance}
\end{equation}

\begin{figure}[t]
\includegraphics[clip,width=0.80\linewidth]{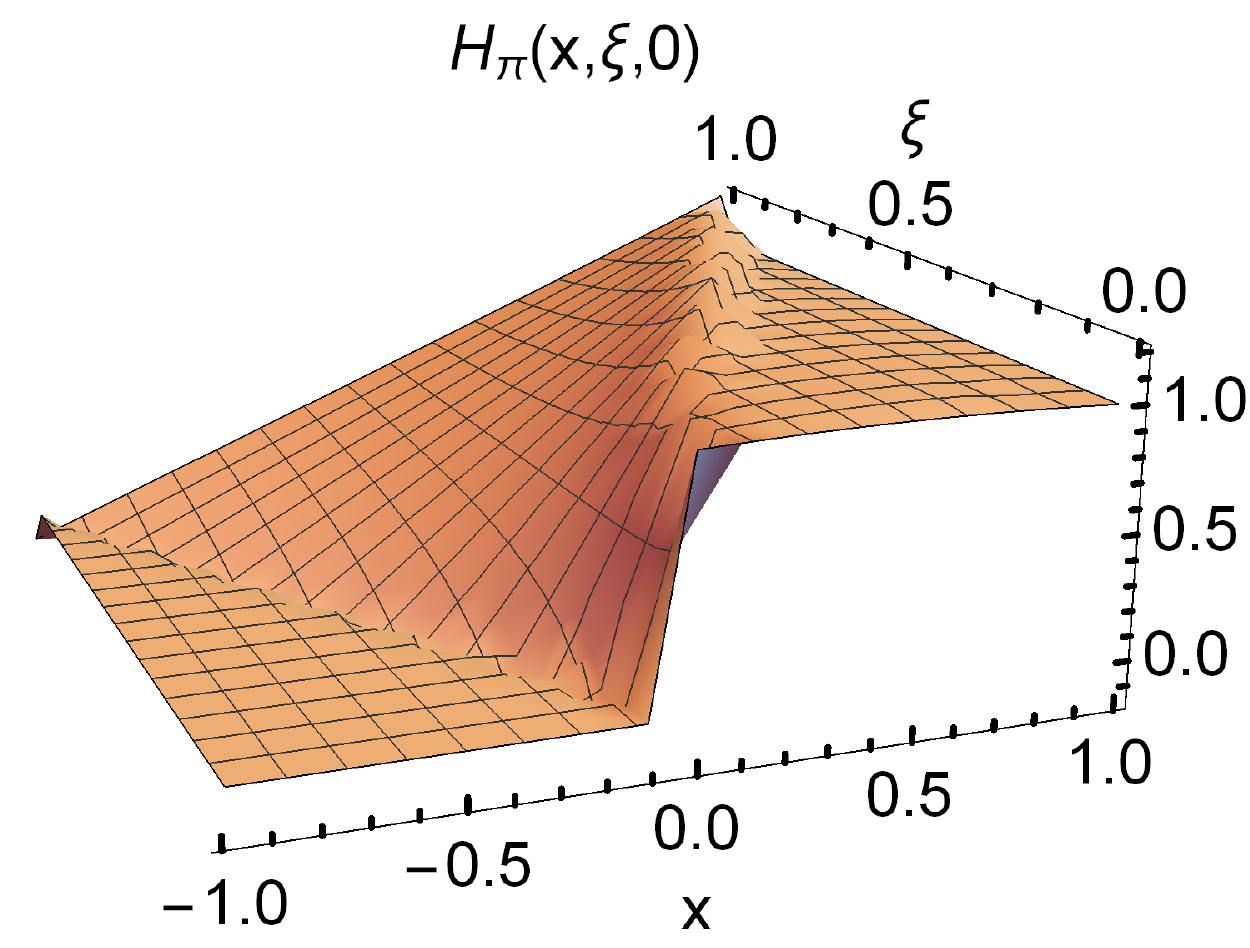}\\[2ex]
\includegraphics[clip,width=0.80\linewidth]{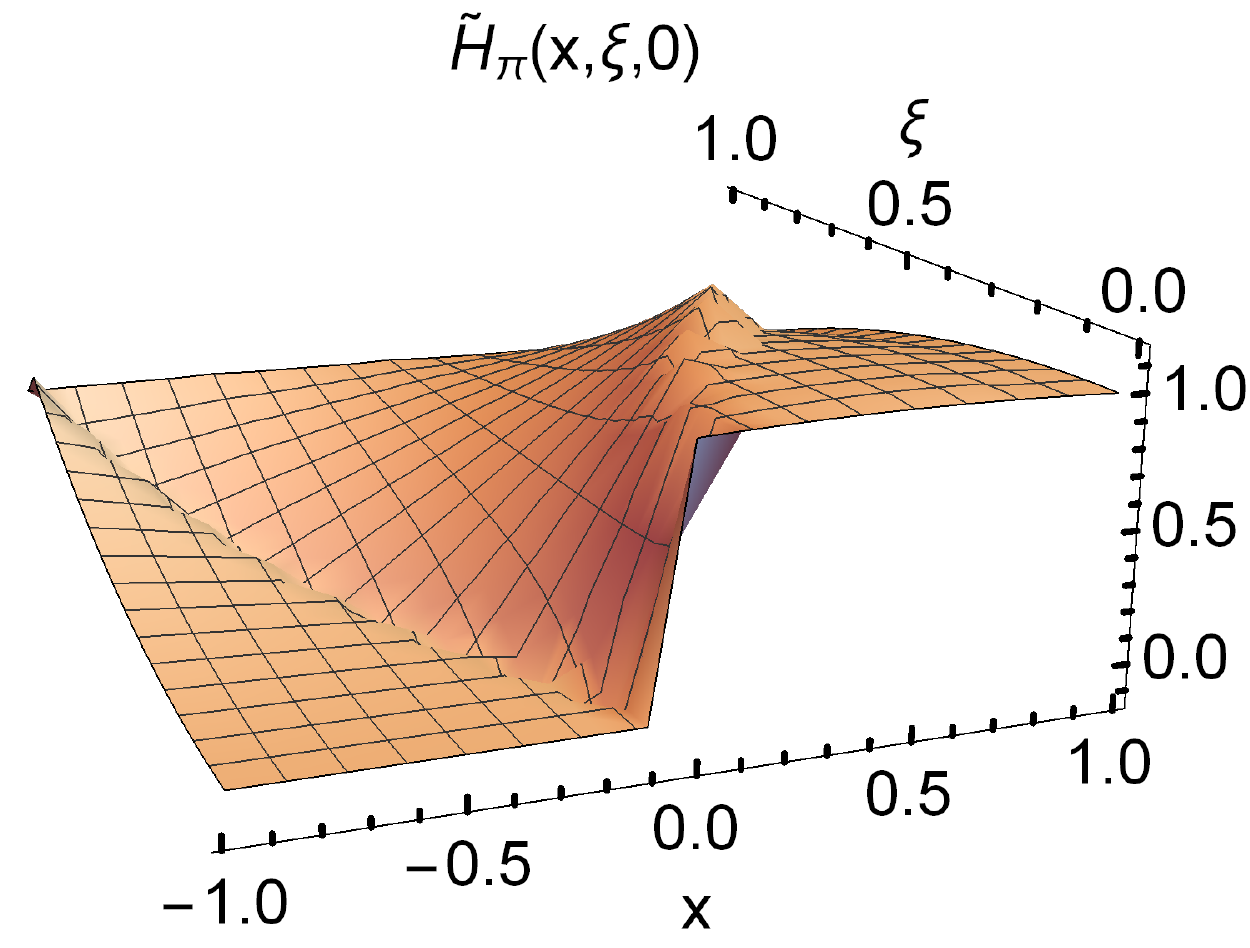}

%
\caption{\label{DrawHpi}
${\mathsf H}_\pi(x,\xi,t=0)$, twist-two vector GPD: \emph{upper panel} -- Eq.\,\eqref{EqGPDFA}; and \emph{lower panel} -- Eq.\,\eqref{EqGPDFA} amended through addition of Eq.\,\eqref{HpiCorrected}.
Owing to Eq.\,\eqref{TRinvariance}, we only plot $\xi>0$.
}
\end{figure}


\subsection{Vector GPD -- Images}
\label{SecVGPDI}
The twist-two vector GPD in Eqs.\,\eqref{EqGPDFA}, \eqref{HpiCorrected} is drawn in Fig.\,\ref{DrawHpi}.
Some features are obvious.
(\emph{a}) ${\mathsf H}_\pi(x, 0, 0)$ is the CI valence-quark parton distribution function, which is $q^{\rm CI}_\pi(x) \approx \theta(x)\theta(1-x)$ at the pion mass in Table~\ref{pionresults}.
(\emph{b}) ${\mathsf H}_\pi(x,\xi,0)= 0$ on $x<-\xi$.  (As we have defined $\tilde{\mathsf H}_\pi$, this is only approximately true; but if necessary, that is readily corrected following the procedure in footnote~\ref{fn1}.)
(\emph{c}) ${\mathsf H}_\pi(x,1,0)= \tfrac{1}{2}\varphi^{\rm CI}_\pi([1+x]/2)$, \emph{i.e}.\ the CI dressed-quark distribution amplitude.
(\emph{d}) Using a contact interaction, the GPD is continuous but not differentiable at $x=\pm \xi$.  (This is typical of models whose basis is a separable interaction \cite{Theussl:2002xp, Broniowski:2007si}.)

Beginning with ${\mathsf H}_\pi$, the pion elastic electromagnetic form factor is obtained via
\begin{equation}
F_\pi^{\rm em}(\Delta^2) = \int_{-1}^{1}dx\, {\mathsf H}_\pi(x,\xi,-\Delta^2)\,.
\end{equation}
It is readily verified by straightforward calculation that the evaluated integral is independent of $\xi$.

\begin{figure}[t]
\includegraphics[clip,width=0.80\linewidth]{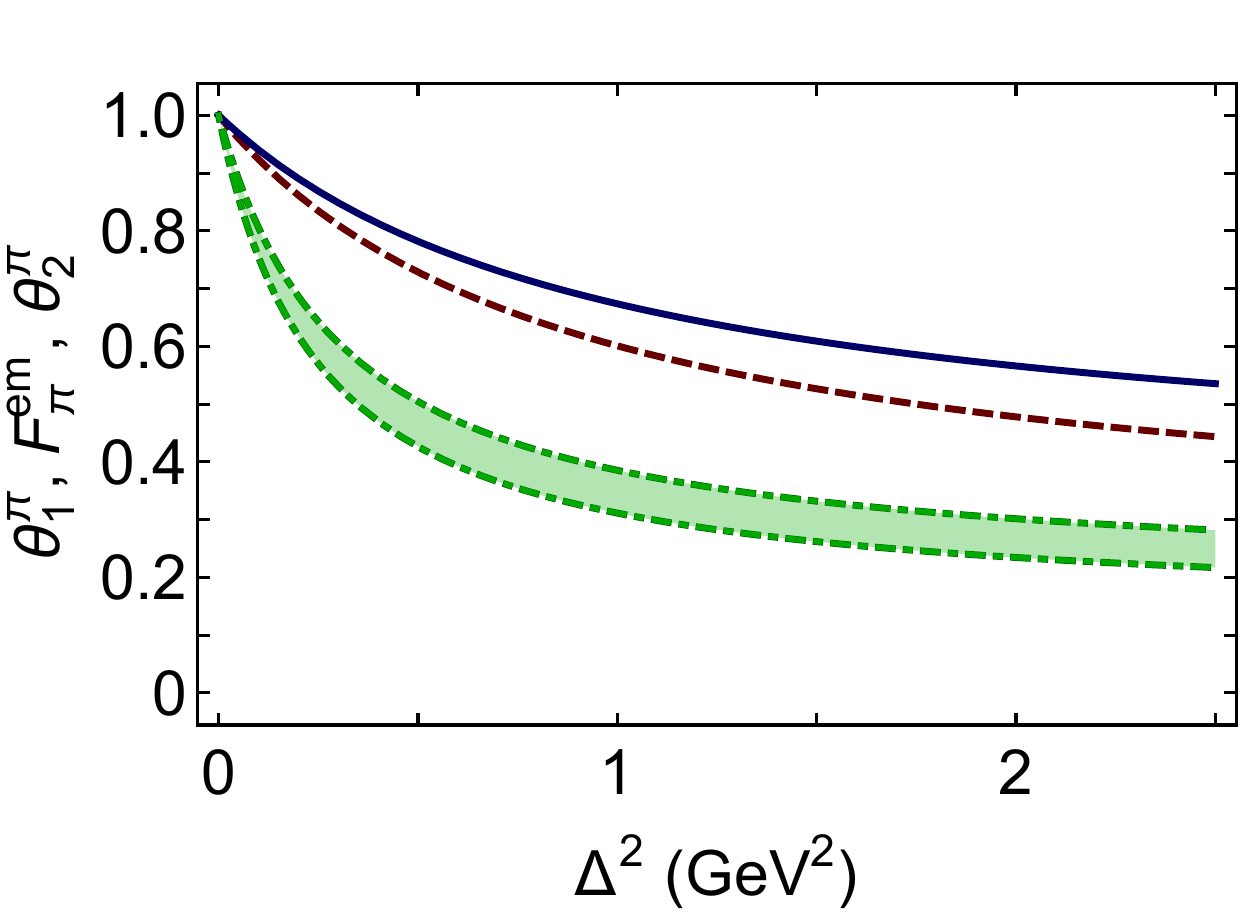}
\includegraphics[clip,width=0.80\linewidth]{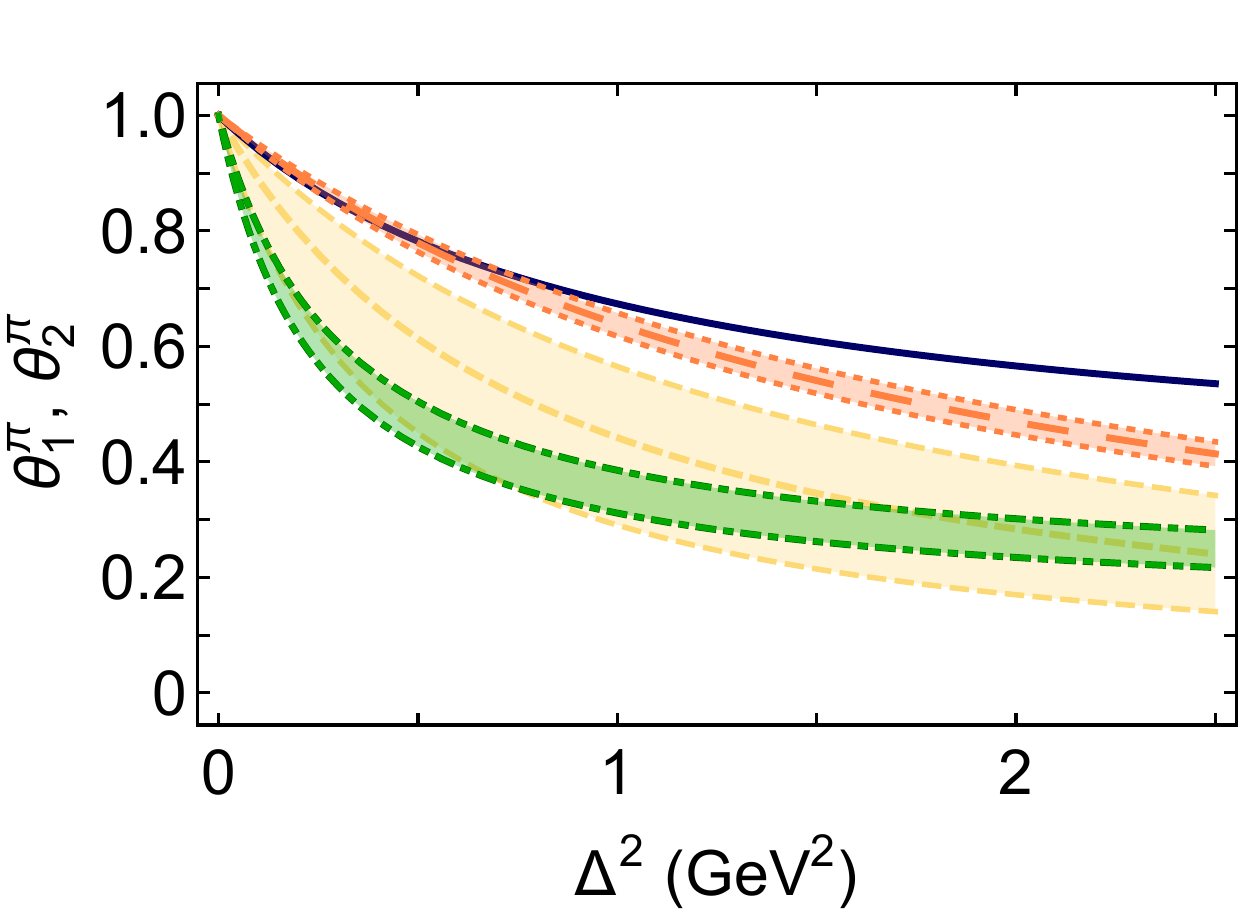}
\caption{\label{DrawFpiQ2}
\emph{Upper panel}.  Pion form factors computed from the twist-two vector GPD:
 solid blue curve -- mass distribution, $\theta_2$;
 dashed red curve -- elastic electromagnetic, $F_\pi^{\rm em}$;
 and green dot-dashed curves with associated band -- pressure, $\theta_1$.
\emph{Lower panel}.  Comparison of CI results for $\theta_{1,2}$ with those obtained using lQCD, \emph{viz}.\ $\theta_1^{\rm lQCD}$ -- yellow short-dashed curve within like-coloured band; and $\theta_2^{\rm lQCD}$ -- orange long-dashed curve and band.  The bands enclose the envelope of curves that fit the lQCD results \cite{Brommel:2007zz}.
}
\end{figure}

The computed pion form factor is depicted in Fig.\,\ref{DrawFpiQ2}\,--\,upper-panel as the dashed red curve, from which one obtains the associated radius: $r_\pi^{\rm em}=0.44\,$fm.  As discussed in detail elsewhere \cite{GutierrezGuerrero:2010md, Roberts:2011wy, Chen:2012txa}, the WGTI-pre\-ser\-ving treatment of a CI necessarily generates $F_\pi \neq 0$ in Eq.\,\eqref{BSAcontactpion}.  Consequently, the CI form factor is hard; namely, it approaches a nonzero constant value as $Q^2\to \infty$.

It is appropriate now to consider the CI pion vector GPD in impact parameter space \cite{Burkardt:2000za}:
\begin{equation}
q_\pi(x,|b_\perp|) = \int_0^\infty\!\frac{ d|\Delta|}{2\pi} \,\Delta\, J_0(|b_\perp ||\Delta|)\,{\mathsf H}_\pi(x,\xi=0,-\Delta^2)\,,
\label{qpixb}
\end{equation}
where $J_0$ is a Bessel function.
This density describes the probability of finding a dressed-quark within the light-front at a transverse position $\vec{b}_\perp$ from the pion's centre of transverse momentum (CoTM).  Inspecting Eqs.\,\eqref{EqGPDFA}\,--\,\eqref{DefintionsHpi} and using Eqs.\,\eqref{Eqsigmas}, it becomes clear that, in contrast to results obtained with realistic interactions \cite{Mezrag:2014jka}, a CI treatment of the pion does not introduce strong $x$-$t$ correlations.  Hence, a fair estimate of $q_\pi(x,b_\perp)$ is obtained by writing ${\mathsf H}_\pi(x,0,-\Delta^2) \approx q_\pi(x) F_\pi^{\rm em}(\Delta^2)$.  Consequently, if one omits $F_\pi$ in Eq.\,\eqref{BSAcontactpion} so that the pion's elastic electromagnetic form factor is a monopole characterised by a length-scale, $r_\pi=\surd 6/M_F$: $F_\pi^{\rm em}(Q^2) = 1/(1+Q^2/M_F^2)$, then
\begin{equation}
q_\pi(x,|b_\perp|) \stackrel{F_\pi = 0}{\approx} q_\pi^{\rm CI}(x) M_F^2 K_0(|b_\perp| M_F)\,,
\end{equation}
where $K_0$ is a modified Bessel function of the second kind.  Returning to an internally consistent WGTI-pre\-ser\-ving CI treatment, so that $F_\pi \neq 0$, then the large-$Q^2$ behaviour of the pion form factor may be characterised via $M_F \to \infty$; hence,
\begin{equation}
q_\pi(x,|b_\perp|) \stackrel{F_\pi \neq 0}{\approx} q_\pi^{\rm CI}(x) \delta^2(\vec{b}_\perp).
\end{equation}
We have verified these statements numerically.

The $n=1$ Mellin moment of the twist-two vector GPD delivers the pion's gravitational form factors:
\begin{equation}
\int_{-1}^1 dx\, 2x \, \tilde {\mathsf H}_\pi(x,\xi,-\Delta^2) = \theta_2^\pi(\Delta^2) - \xi^2 \, \theta_1^\pi(\Delta^2)\,,
\end{equation}
where $\theta_2$ relates to the quark mass distribution within the pion and $\theta_1$ is linked to the quark pressure distribution.  In a symmetry preserving treatment: $\theta_2^\pi(0) = 1$; and, following from Eqs.\,\eqref{EqSoft}, $\theta_1^\pi(0) - \theta_2^\pi(0) = {\rm O}(m_\pi^2)$.\footnote{Recall Observation\,\emph{B} following Eq.\,\eqref{EqKinematics}; to wit, the results presented here are defined at the hadronic scale, $\zeta_H$, whereat all properties of the bound-state are invested in the dressed-quark and dressed-antiquark quasiparticles.}

The pion's gravitational form factors are also drawn in Fig.\,\ref{DrawFpiQ2}.
Regarding $\theta_1^\pi$, the \emph{Ansatz} used to correct $ {\mathsf H}_\pi$ on the domain ${\mathpzc E}$, Eqs.\,\eqref{HpiCorrected}, depends on a representation of the $\sigma$-resonance contribution to quark+quark scattering in the scalar channel.  To illustrate the associated model-dependent uncertainty, we used two forms:
\begin{subequations}
\begin{align}
P_\sigma^{\rm CI}(t) & = 1/(1-t/[4 M^2])\,, \\
P_\sigma^{\rm emp}(t) & = 1/|1-t/m_{f_0}^2|\,,
\end{align}
\end{subequations}
where $m_{f_0} \approx (0.48 - i 0.28)\,$GeV \cite{Zyla:2020}.  The first choice is based on the observation that the CI produces a $\sigma$-meson with mass $m_\sigma \approx 2 M$ in the neighbourhood of the chiral limit \cite{Roberts:2011wy}, whereas the second uses instead the pole mass associated with the empirical $\sigma$-resonance.  Evidently, the uncertainty is noticeable but not large.  We find $r_{\theta_1}^\pi/r_\pi^{\rm em} = 1.88(13)$; and a result that is generally softer than the pion's electromagnetic form factor.
Turning to $\theta_2^\pi$, $r_{\theta_2}^\pi/r_\pi^{\rm em} = 0.89$; and this form factor is generally harder than $F_\pi^{\rm em}(\Delta^2)$.

The lower panel of Fig.\,\ref{DrawFpiQ2} displays a comparison between our CI results and those obtained using lattice-QCD (lQCD), described by \cite{Brommel:2007zz}:
\begin{equation}
\label{EqlQCDtheta}
\theta^{{\pi}_{\rm lQCD}}_{1,2}(\Delta^2) = 1/[1 + \Delta^2/M_{1,2}^2]\,,
\end{equation}
$M_1 = 0.89 (25)\,$GeV, $M_2 = 1.33 (2)\,$GeV.  The errors on $M_{1,2}$ lead to bands which demarcate the envelope of curves that provide a reasonable fit to the actual (scattered) lQCD results.  Evidently, there is fair semiquantitative agreement between the CI and lQCD results, especially allowing for the hardness of CI form factors.

Working with such hadron form factors, Ref.\,\cite{Polyakov:2018zvc} defined Breit-frame pressure distributions, \emph{e.g}.\
\begin{subequations}
\label{EqPressure}
\begin{align}
p_\pi&(r)  = \frac{1}{3} \int \frac{d^3\vec{\Delta}}{(2\pi)^3}\frac{1}{2E(\Delta)}\, {\rm e}^{i\vec{\Delta}\cdot \vec{r}}\, [\Delta^2\theta_1^\pi(\Delta^2)] \\
& = \frac{1}{6\pi^2 r} \int_0^\infty d\Delta \,\frac{\Delta}{2 E(\Delta)} \, \sin (\Delta r) [\Delta^2\theta_1^\pi(\Delta^2)] \,,
\end{align}
\end{subequations}
where $2E(\Delta)=\sqrt{4 m_\pi^2+\Delta^2}$.
The physical interpretation of such distributions is complicated by issues connected with the Poincar\'e transformation of frame-dependent wave functions in quantum field theory \cite{Miller:2010nz}.  Nevertheless, they are mathematically well defined; do admit the standard interpretation in systems for which a nonrelativistic approximation can be discussed; and viewed judiciously, can deliver fruitful insights.
Moreover, two-dimensional Fourier-transform analogues deliver results of similar magnitude.

\begin{figure}[t]
\includegraphics[clip,width=0.82\linewidth]{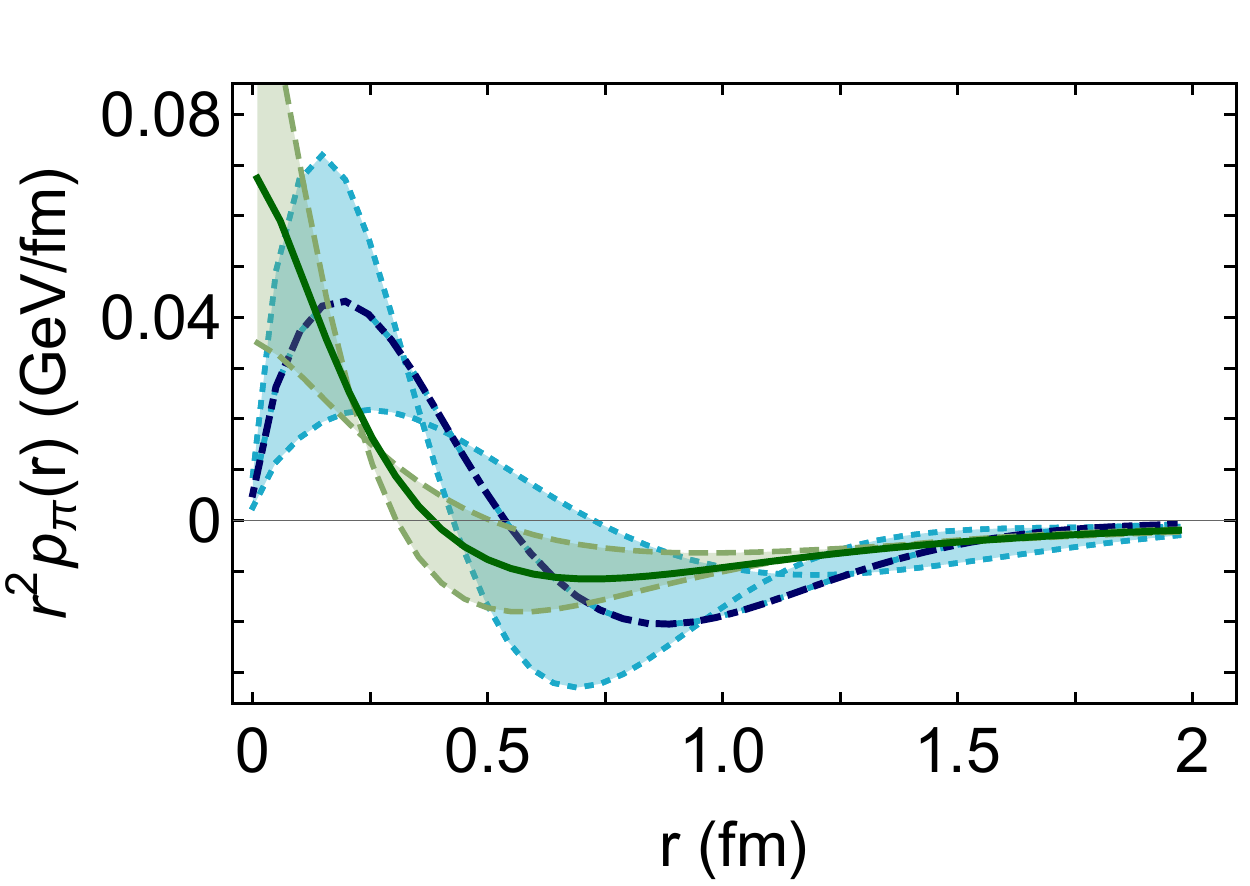}
\includegraphics[clip,width=0.82\linewidth]{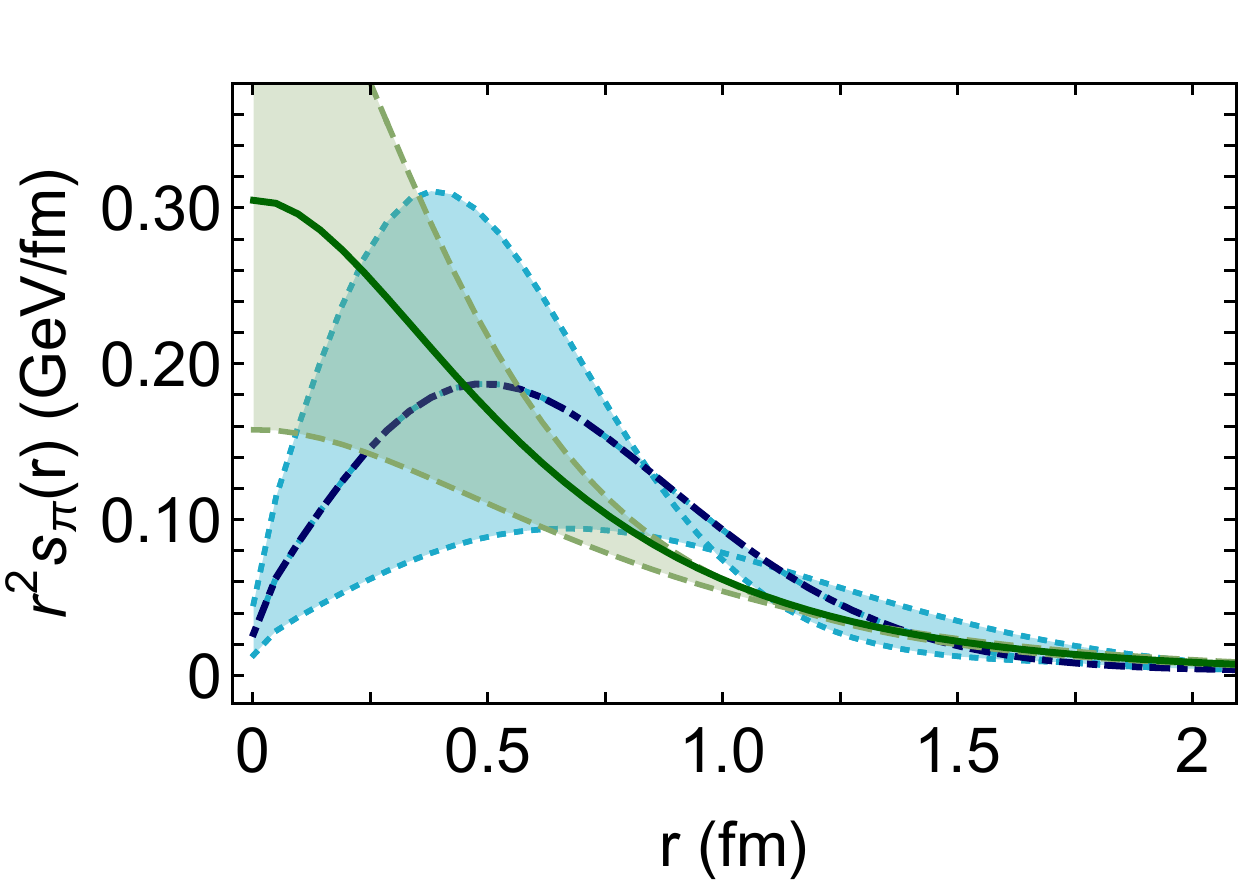}
\caption{\label{Drawpressurepi}
\emph{Upper panel} -- pressure distribution in the pion, Eq.\,\eqref{EqPressure}; and \emph{lower panel} shear pressure distribution, Eq.\,\eqref{EqShear}.
Legend.
Green solid curve within like-coloured band -- computed using the lQCD results for $\theta_1(\Delta^2)$ in Eq.\,\eqref{EqlQCDtheta}; and
blue dot-dashed curve and associated band -- computed using $\theta_1(\Delta^2)$ in Eq.\,\eqref{EqlQCDthetaSV}.
}
\end{figure}

Owing to the hardness of CI pion form factors, the integrals in Eqs.\,\eqref{EqPressure} do not converge when evaluated using the results for $\theta_{1,2}^\pi(\Delta^2)$ depicted in Fig.\,\ref{DrawFpiQ2}\,--\,upper panel.  We therefore exploit the semiquantitative similarity between CI and lQCD results evident Fig.\,\ref{DrawFpiQ2}\,--\,lower panel to justify an estimate of the pion's pressure distribution using Eq.\,\eqref{EqlQCDtheta}.  The result is depicted in Fig.\,\ref{Drawpressurepi} and the qualitative features are consistent with an intuitive physical interpretation.  Namely, the pressure is large and positive in the neighbourhood $r\simeq 0$ -- the dressed-quark+dressed-antiquark are pushing away from each other at small separation; but the pressure changes sign as the separation becomes large, signalling a transition into the domain whereupon the pair experience the effects of confinement forces.

It is important to appreciate that $ \lim _{r\to 0}r^2 p_\pi(r) \neq 0$ in Fig.\,\eqref{Drawpressurepi} is an artefact of the simple monopole description of $\theta_1(\Delta^2)$ in Eq.\,\eqref{EqlQCDtheta}.  In four spacetime dimensions, a quantum field theoretical treatment of form factors always introduces scaling violations, leading to additional $\ln (\Delta^2/M^2)$ suppression on $\Delta^2\gg M^2$.  We choose to illustrate the effect of such scaling violation by modifying Eq.\,\eqref{EqlQCDtheta} as follows:
\begin{equation}
\label{EqlQCDthetaSV}
\theta^{{\pi}_{\rm lQCD}}(y=\Delta^2/M^2) = 1/[1 + y\ln(1+y)]\,.
\end{equation}
Using this form for $\theta_1$ leads to the blue dot-dashed curve in Fig.\,\eqref{Drawpressurepi}.  In this case, $\lim _{r\to 0}r^2 p_\pi(r) = 0$; yet, the characterising magnitudes are unchanged.

An analogue of Eq.\,\eqref{EqPressure} has been used to infer the proton's quark pressure distribution from existing data on deeply virtual Compton scattering \cite{Burkert:2018bqq}.  Comparing that result with those in Fig.\,\ref{Drawpressurepi}\,--\,upper panel, one observes that: (\emph{i}) the pressure within the pion on the neighbourhood $r\simeq 0$ is roughly five-times larger than that in the proton; and (\emph{ii}) the two pressure profiles have a similar radial extent.
Notwithstanding the issues with Ref.\,\cite{Burkert:2018bqq} canvassed in Refs.\,\cite{Kumericki:2019ddg, Moutarde:2019tqa}, profiles analogous to Fig.\,\ref{Drawpressurepi}\,--\,upper panel for neutron stars indicate $r\simeq  0$ pressures therein of roughly $0.1\,$GeV/fm \cite{Ozel:2016oaf}; hence, the core pressures in the pion and neutron stars are commensurate.

A shear pressure distribution can also be defined \cite{Polyakov:2018zvc}:
\begin{subequations}
\label{EqShear}
\begin{align}
s_\pi&(r)  = -\frac{3}{4} \int \frac{d^3\vec{\Delta}}{(2\pi)^3}\frac{ {\rm e}^{i\vec{\Delta}\cdot \vec{r}}\,}{2E(\Delta)}\, P_2(\hat{\Delta}\cdot \hat{r}) [\Delta^2\theta_1^\pi(\Delta^2)] \\
& = \frac{3}{16\pi^2} \int_0^\infty d\Delta \,\frac{\Delta}{2 E(\Delta)} \,\Delta\, {\mathpzc j}_2 (\Delta r) [\Delta^2\theta_1^\pi(\Delta^2)] \,,
\end{align}
\end{subequations}
where $\hat \Delta^2=1=\hat r^2$ and ${\mathpzc j}_2$ is a spherical Bessel function.  Intuitively, $r^2 s_\pi(r)$ provides an indication of the strength of QCD forces within the pion which act to deform it.  Our results are drawn in Fig.\,\ref{Drawpressurepi}\,--\,lower panel.  Focusing on the more realistic curve, obtained using Eq.\,\eqref{EqlQCDthetaSV}, these forces peak in the neighbourhood upon which the normal pressure switches sign, \emph{i.e}.\ where the forces driving the quark and antiquark away from the core are overwhelmed by attractive confinement pressure.

\begin{figure}[t]
\includegraphics[clip,width=0.80\linewidth]{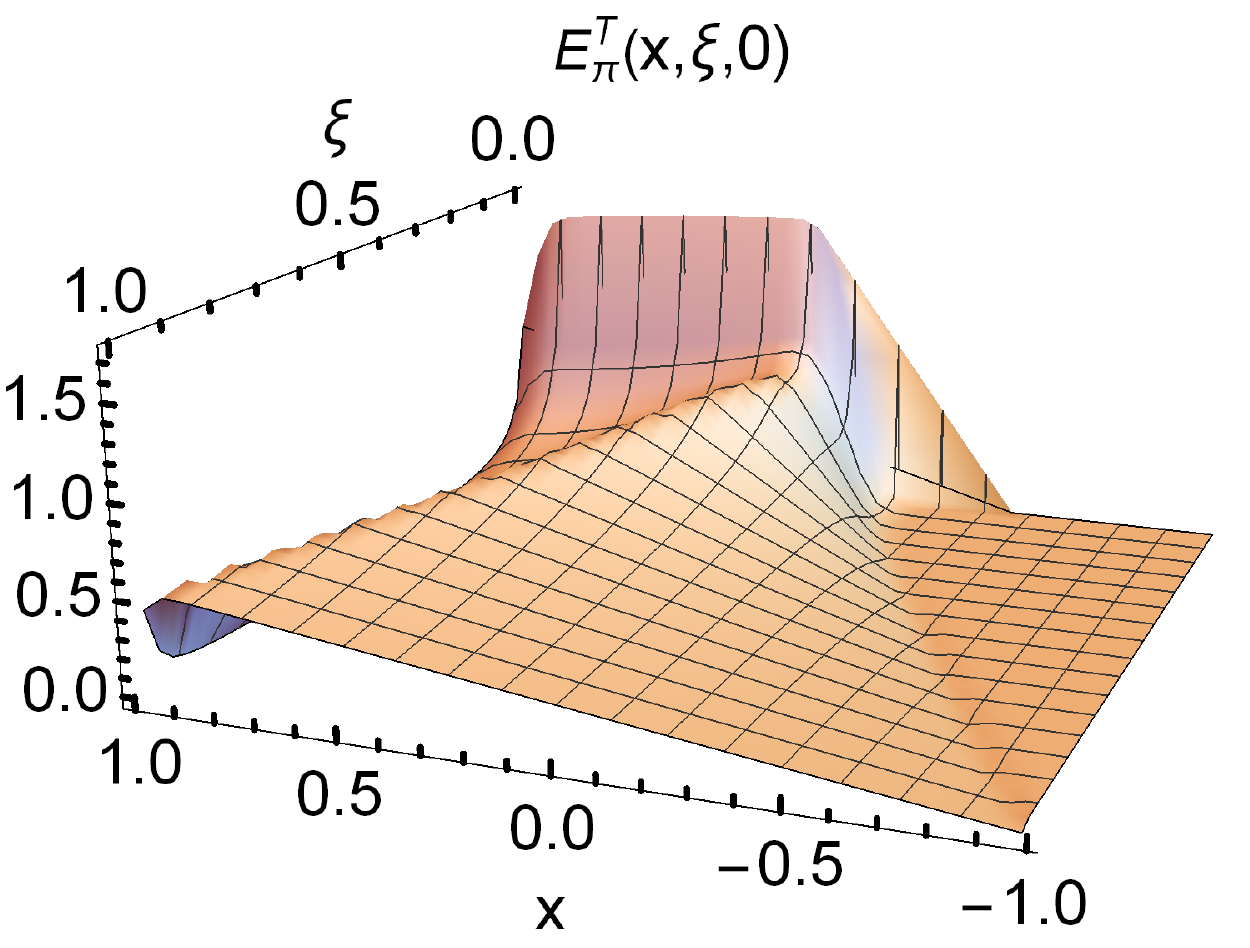}
\caption{\label{DrawEpi}
${\mathsf E}_\pi^{\rm T}(x,\xi,t=0)$ -- twist-two tensor GPD, Eq.\,\eqref{EqEpi}.  Owing to Eq.\,\eqref{EqEpTRinvariance}, only $\xi>0$ is plotted.
}
\end{figure}

\subsection{Tensor GPD -- Images}
\label{SecTensorGPD}
The twist-two tensor GPD expressed in Eqs.\,\eqref{EqEpi} is drawn in Fig.\,\ref{DrawEpi}: it is only nonzero on $-\xi< x<1$.  Working with this distribution, one obtains the following tensor form factors as the leading Mellin moments:
{\allowdisplaybreaks
\begin{subequations}
\label{EqB1B2}
\begin{align}
{B}^\pi_{10} &(-\Delta^2) = \int_{-1}^1 dx \, {\mathsf E}_\pi^{\rm T}(x,\xi,-\Delta^2) \\
& =  P_T(\Delta^2) \frac{N_c}{4\pi^2}\left[
-\frac{F_\pi^2}{M} \int_0^1 dx\, \bar{\mathpzc C}_1(\sigma_5) \right. \nonumber \\
& \left.
\quad + 2 N_{EF}
\int_0^1 dx \!\int_{0}^{1-x}\! dy\, \frac{M}{\sigma_6}\bar{\mathpzc C}_2(\sigma_6)
\right] , \label{EqB10} \\
{B}^\pi_{20} &(-\Delta^2) = \int_{-1}^1 dx \, x\,{\mathsf E}_\pi^{\rm T}(x,0,-\Delta^2) \\
& =P_T(\Delta^2)  \frac{N_c}{2\pi^2} N_{EF} \nonumber \\
& \quad \times  \int_0^1 dx\! \int_0^{1-x}\! dy\, (1-x-y) \frac{M}{\sigma_6}\bar{\mathpzc C}_2(\sigma_6)\,.
\end{align}
\end{subequations}}

Evaluated using the CI parameters in Table~\ref{pionresults},
\begin{subequations}
\label{EqB1B2zero}
\begin{align}
M\,B^\pi_{10}(0) & = 0.18\,, & M\,B^\pi_{20}(0) = 0.070\,,\; \\
m_\pi\,B^\pi_{10}(0) & = 0.070\,, &  m_\pi\,B^\pi_{20}(0) = 0.026\,,\; \\
B^\pi_{10}(0) /B^\pi_{20}(0) & = 2.65\,. &
\end{align}
\end{subequations}
These quantities are subject to QCD evolution; and, as described after Eq.\,\eqref{EqKinematics}, we interpret the results in Eq.\,\eqref{EqB1B2zero} as being valid at the hadronic scale, the value of which is discussed in Refs.\,\cite{Cui:2020dlm, Cui:2020piK}:
\begin{equation}
\label{HadronScale}
\zeta_H=0.331(2)\,{\rm GeV}.
\end{equation}

Using QCD's infrared-finite process-independent effective charge \cite{Cui:2019dwv}, $\hat\alpha(k^2)$, to integrate the evolution equations \cite{Cui:2020dlm, Cui:2020piK}, one finds
\begin{equation}
B_{n0}(0;\zeta_F) = B_{n0}(0;\zeta_H)
\exp\left[
\frac{\gamma_{0(n)}^{qq{\rm T}}}{4\pi}
\int_{t_F}^{t_H}\, dt\, \hat\alpha({\rm e}^{t})
\right]\,, \label{Bn0Evolution}
\end{equation}
where $t_F = \ln\zeta_F^2$, $t_H=\ln\zeta_H^2$ and \cite[Eq.\,(4.160)]{Belitsky:2005qn}
\begin{equation}
\gamma_{0(n)}^{qq{\rm T}} = -\frac{4}{3}\left[ 3 - 4 \sum_{k=1}^{n+1}\frac{1}{k}\right].
\end{equation}
Consequently, at $\zeta=\zeta_2=2\,$GeV,
\begin{subequations}
\label{EqB1B2zeta2}
\begin{align}
m_\pi\,B^\pi_{10}(0) & = 0.053\,, &  m_\pi\,B^\pi_{20}(0) = 0.012\,,\; \\
B^\pi_{10}(0) /B^\pi_{20}(0) & = 4.57\,. &
\end{align}
\end{subequations}
This is the scale used in Ref.\,\cite{Brommel:2007xd}, which reports the following values for these quantities after an extrapolation to the physical pion mass: $0.22(3)$, $0.039(10)$, $5.66(60)$, in qualitative agreement with the CI results.  Similar conclusions are drawn elsewhere, \emph{e.g}.\ Refs.\,\cite{Nam:2010pt, Dorokhov:2011ew, Fanelli:2016aqc}.

\begin{figure}[t]
\includegraphics[clip,width=0.80\linewidth]{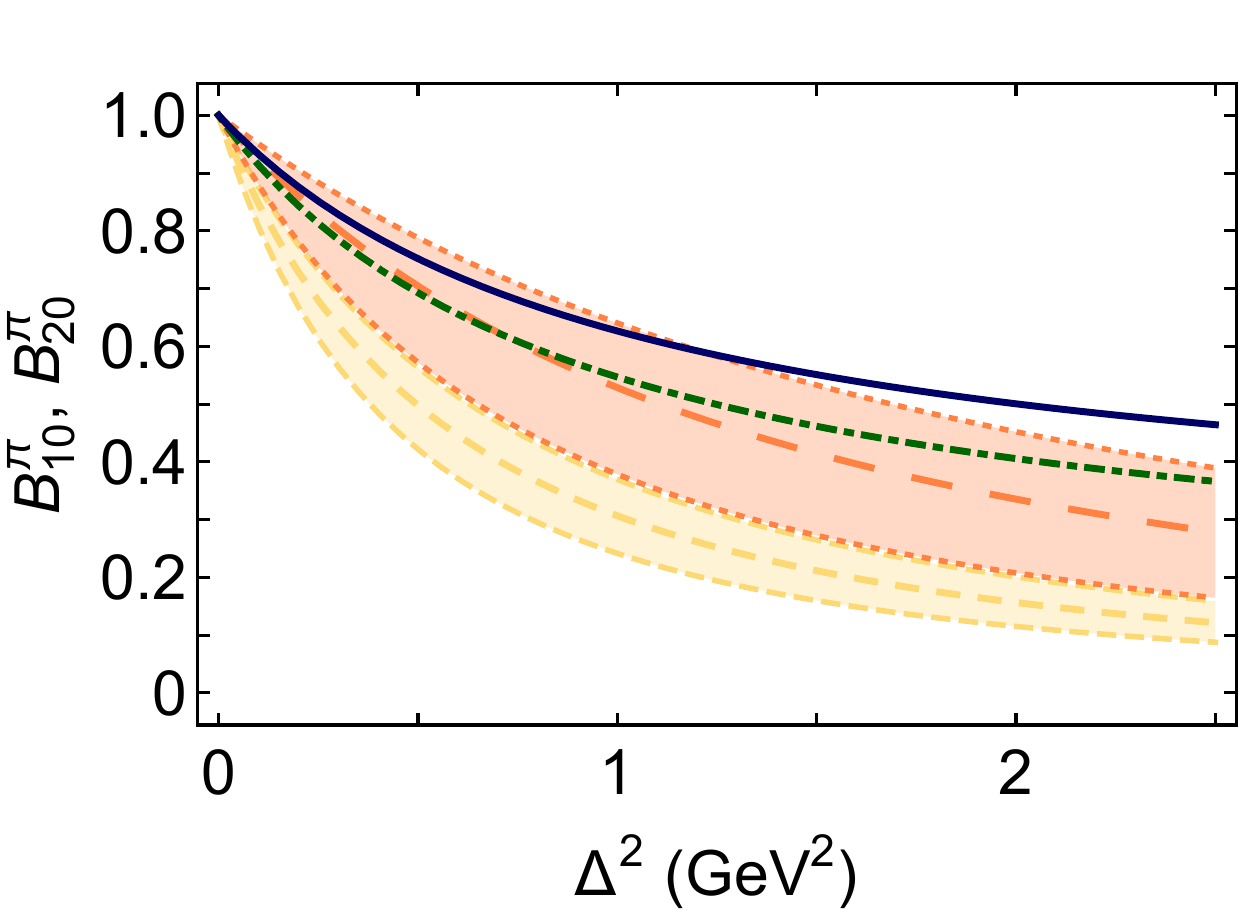}
\caption{\label{FigB1B2}
Twist-two tensor form factors, Eq.\,\eqref{EqB1B2}, normalised to unity at $t=0$ using the results in Eq.\,\eqref{EqB1B2zero}:
dot-dashed green curve -- ${B}^\pi_{10}$; and  solid blue curve -- ${B}^\pi_{20}$.
Normalised this way, the depicted form factors are independent of the renormalisation scale.
For comparison, lQCD results from Ref.\,\cite{Brommel:2007xd}:
short-dashed yellow curve within like-coloured band -- ${B}^\pi_{10}$ and
long-dashed orange curve and band -- ${B}^\pi_{20}$.
As in Fig.\,\ref{DrawFpiQ2}, the bands enclose the envelope of curves that fit the lQCD results.
}
\end{figure}

The tensor form factors in Eqs.\,\eqref{EqB1B2} are plotted in Fig.\,\ref{FigB1B2}, normalised by their $\Delta^2=0$ values.  Employing this procedure, the depicted form factors are independent of the renormalisation scale \cite{Fanelli:2016aqc}.  Hence, comparison with the lQCD results in Ref.\,\cite{Brommel:2007xd} is meaningful, although quantitative agreement cannot be expected because the lQCD form factors were computed using $m_\pi^2 \approx 20 \, m_\pi^{2\,{\rm empirical}}$.  Bearing this in mind and considering that the CI produces hard pseudoscalar meson form factors, there is reasonable qualitative agreement, \emph{e.g}.: the radii have the same ordering, $r_{B1}^\pi/r_{B_{20}^\pi} = 1.48(17)$ (lQCD) vs.\ 1.14 (herein); and $B_{10}(t)$ is generally softer than $B_{20}(t)$.

One now has access to the light-front transverse-spin distribution of dressed-quarks within the pion, which is defined in impact-parameter space \cite{Brommel:2007xd}:
\begin{equation}
\label{EqSpinDensity}
\rho_1(b_\perp,s_\perp) = \tfrac{1}{2}\tilde q_\pi(|b_\perp|) - \tfrac{1}{2}\varepsilon^{ij} s_\perp^i b_\perp^j B^{\prime\pi}_{10}(|b_\perp|)\,,
\end{equation}
with
\begin{subequations}
\begin{align}
\tilde q_\pi(|b_\perp|) & = \int_{-1}^1dx\, q_\pi(x,|b_\perp|)\,,\\
B^{\prime\pi}_{10}(|b_\perp|) & = -\frac{1}{4\pi |b_\perp|}
\int_0^\infty \! d |\Delta|\, \Delta^2\, J_1(|b_\perp||\Delta|) B_{10}^\pi(-\Delta^2)\,,
\end{align}
\end{subequations}
where $q_\pi(x,|b_\perp|)$ is given in Eq.\,\eqref{qpixb} and $J_1$ is a Bessel function.  For a dressed-quark polarised in the $+x$ direction and $\hat s_\perp\cdot \hat b_\perp= \cos\phi_\perp$, $\varepsilon^{ij} s_\perp^i b_\perp^j = |b_\perp| \sin\phi_\perp$.

\begin{figure}[t]
\includegraphics[clip,width=0.90\linewidth]{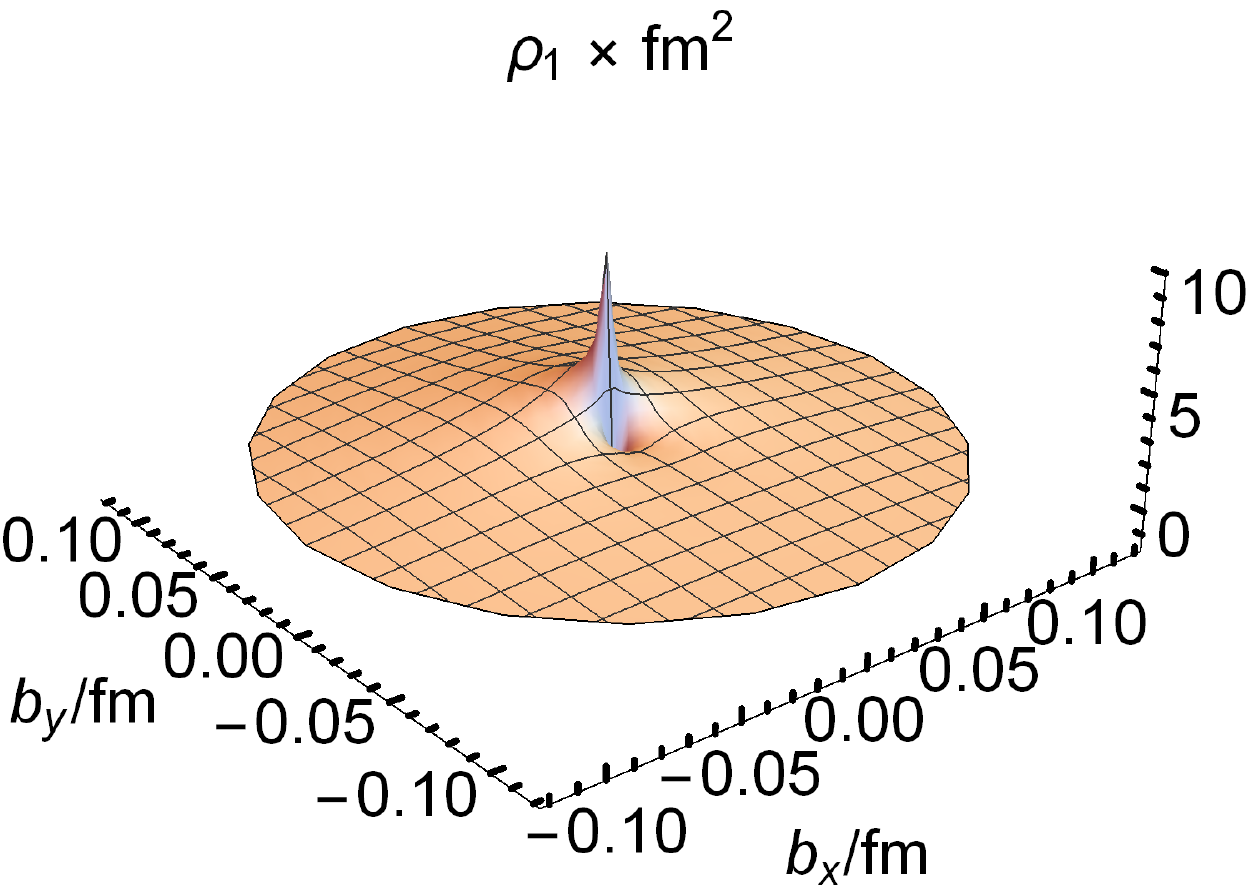}
\includegraphics[clip,width=0.80\linewidth]{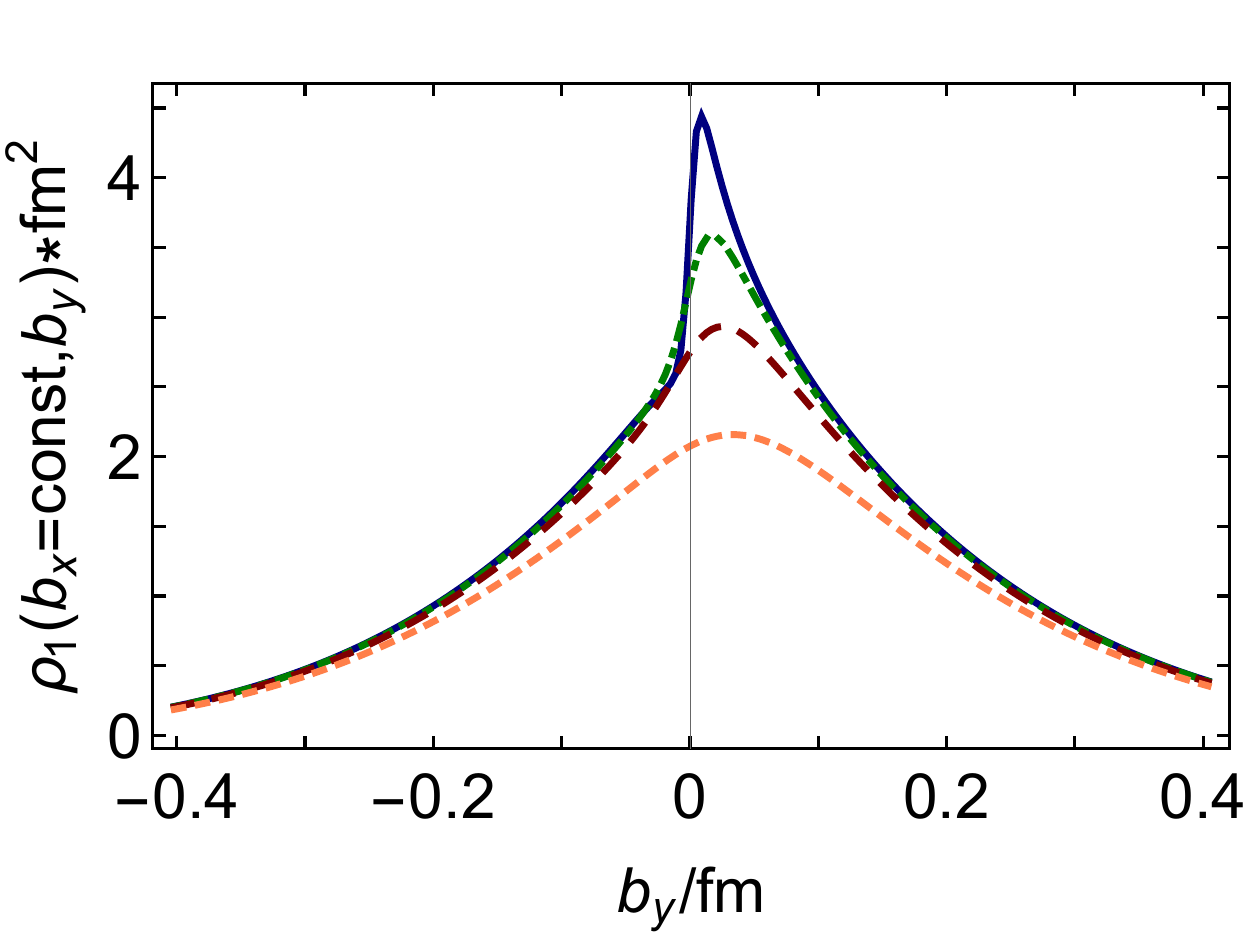}

%
\caption{\label{Drawrho1pi}
$\rho_1(b_\perp,s_\perp \propto \hat x)$, Eq.\,\eqref{EqSpinDensity}, light-front transverse-spin distribution of dressed valence quarks within the pion.  \emph{Upper panel} -- full three-dimensional image; and \emph{lower panel} -- slices at constant $b_x/$fm: solid blue -- $b_x=0.01$; dot-dashed green -- $b_x=0.025$; dashed red -- $b_x=0.05$; and short-dashed orange -- $b_x=0.1$.
In both panels, the scale is $\zeta_H$, Eq.\,\eqref{HadronScale}.
}
\end{figure}

As emphasised above, in an internally consistent CI treatment, all pion form factors are hard; so the integrals that define the transverse densities in Eq.\,\eqref{EqSpinDensity} are ill defined.  It is nevertheless worth illustrating the character of $\rho_1(b_\perp,s_\perp)$.  We therefore employ the expedient introduced in Eq.\,\eqref{EqlQCDthetaSV}, choosing the mass-scale ``M'' to reproduce the CI result for the $\Delta^2\simeq 0$ slope of a monopole approximation to the given form factor and setting its $\Delta^2= 0$ value to match the CI value; to wit,
\begin{subequations}
\begin{align}
F_\pi^{\rm em}(\Delta^2) & = 1/(1+\Delta^2/M_F^2 \ln(1+\Delta^2/M_F^2))\,, \\
B_{10}^\pi(-\Delta^2) & =  (0.070/m_\pi)/(1+\Delta^2/M_B^2 \ln(1+\Delta^2/M_B^2)) \,,
\end{align}
\end{subequations}
with $M_F=1.09\,$GeV, $M_B=1.02\,$GeV.
The result is drawn in Fig.\,\ref{Drawrho1pi}.

Figure\,\ref{Drawrho1pi} shows that for a dressed valence-quark polarised in the light-front-transverse $+x$ direction, the transverse-spin density is no longer symmetric around $\vec{b}_\perp=(b_x=0,b_y=0)$.  Instead, the peak is shifted to $(b_x=0,b_y>0)$, with strength transferred from $b_y<0$ to $b_y>0$.  The average transverse shift is  \cite{Brommel:2007xd}:
\begin{equation}
\langle b_y \rangle = \frac{1}{2} B_{10}(0)/m_\pi = 0.049\,{\rm fm};
\end{equation}
and the $b_y$ profile remains symmetric around the line $b_x=0$.  We interpret these results as being valid at $\zeta_H$.  The distortion vanishes logarithmically with $B_{10}^\pi(0) \to 0$ under QCD evolution, Eq.\,\eqref{Bn0Evolution}.

Given that ${\mathsf E}_{\pi^+}^{{\rm T}\,\bar d}(x,\xi,t) = -{\mathsf E}_{\pi^+}^{{\rm T}\,u}(-x,\xi,t)$, then the three-dimensional profile for a $s_\perp\parallel \hat x$ dressed valence-antiquark is obtained by rotating Fig.\,\ref{Drawrho1pi}\,--\,upper panel by $180^\circ$ around the $b_y=0$ axis.  Regarding Fig.\,\ref{Drawrho1pi}\,--\,lower panel, $b_y\to -b_y$ and the curves change sign.

\begin{figure}[t]
\includegraphics[clip,width=0.80\linewidth]{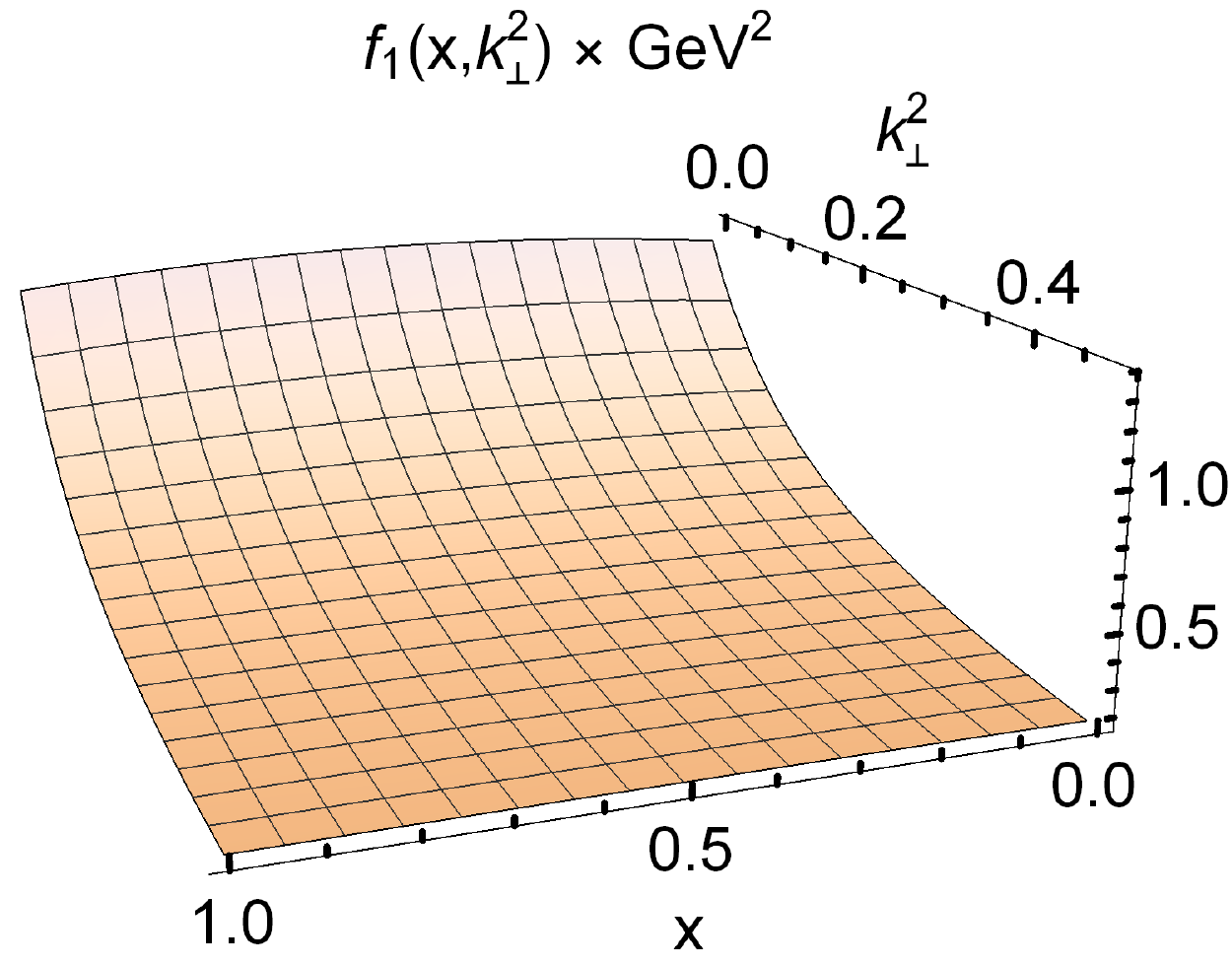}
\includegraphics[clip,width=0.80\linewidth]{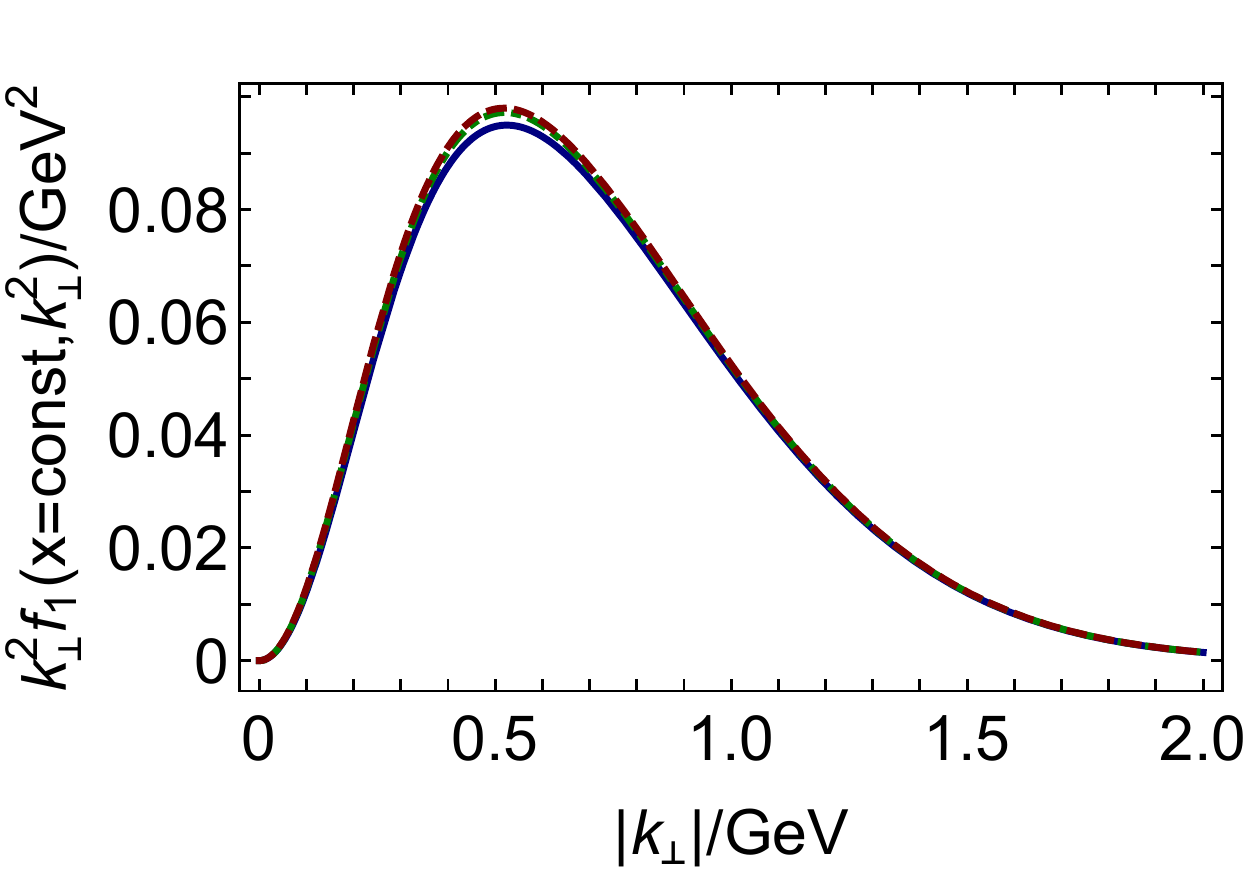}

%
\caption{\label{FigT2TMD}
\emph{Upper panel}.  Twist-two pion TMD, Eq.\,\eqref{EqpionTMDT2}.
This function is symmetric around the line $x=1/2$.
\emph{Lower panel}.  $k_\perp^2 f_1(x,k_\perp^2)$ at: $x=0$ -- solid blue curve; $x=1/4$ -- dot-dashed green curve; and $x=1/2$ -- dashed red curve.
Scale is $\zeta_H$, Eq.\,\eqref{HadronScale}.
}
\end{figure}

\section{Pion TMDs}
\label{SecTMDs}
\subsection{Twist-two TMDs}
Recall now that one proceeds from a given GTMD to the associated TMD by setting $\Delta = 0$, which also means $\xi=0$.  At twist-two, our CI treatment (which does not include a Wilson line) produces one nonzero TMD, whose form can be read from Eq.\,\eqref{EqGTMDF1} ($\varsigma:= \sigma_1^{k_\perp^2,0}$):
\begin{subequations}
\label{EqpionTMDT2}
\begin{align}
f_1(x,k_\perp^2)& = F_1(x,k_\perp^2,0,0) \\
& = \frac{N_c}{2\pi^3}\left[ E_\pi[E_\pi - 2 F_\pi]
\frac{\bar{\mathpzc C}_2(\varsigma)}{\varsigma} \right. \nonumber \\
& \left.
\qquad + 3 \, N_{EF}\,x(1-x) m_\pi^2
\frac{\bar{\mathpzc C}_3(\varsigma)}{\varsigma^2}\right] \,.
\end{align}
\end{subequations}
This TMD, describing the dressed valence $u$-quark in the $\pi^+$, is depicted in Fig.\,\ref{FigT2TMD}.  (Note that $M^2 f_1(x,k_\perp^2) $ is dimensionless.)  The root-mean-square value of $k_\perp^2$ is defined via
\begin{subequations}
\begin{align}
& \langle k_\perp^2\rangle = \int_0^1dx\,\int d^2k_\perp \, k_\perp^2 f_1(x,k_\perp^2) \\
&\Rightarrow \langle k_\perp^2\rangle^{1/2} = 0.61\,{\rm GeV}.
\end{align}
\end{subequations}
Evidently and unsurprisingly, the symmetry-preserving CI-treatment produces a hard $k_\perp^2$ distribution even at the hadronic sale, $\zeta_H$.  In contrast, a pion twist-two TMD developed from an interaction with QCD-like momentum dependence yields  \cite{Xu:2018eii} $\langle k_\perp^2\rangle^{1/2} = 0.21\,$GeV.

Owing to gluon radiation and additional fragmentation, the distributions in Fig.\,\ref{FigT2TMD} become broader as the scale is evolved to values $\zeta>\zeta_H$ \cite{Aybat:2011zv}, whilst nevertheless preserving the result
\begin{equation}
\int d^2k_\perp\, f_1(x,k_\perp^2;\zeta) = u_\pi(x;\zeta)\,,
\end{equation}
which is the $\pi^+$ valence $u$-quark distribution function.

Since we omit the Wilson line, our result for the pion's Boer-Mulders function is
\begin{equation}
h_1^\perp(x,k_\perp^2) \equiv 0\,.
\end{equation}

\subsection{Twist-three TMDs}
In the absence of a Wilson line, the CI supports two nonzero twist-three pion TMDs.  The first is obtained from the GTMD $E_2(x,k_\perp^2,\xi,t)$ in Sec.\,\ref{AppE2}:
\begin{subequations}
\label{EqpionTMDT3e}
\begin{align}
& e(x,k_\perp^2)  = E_2(x,k_\perp^2,0,0)=: \hat e(x,k_\perp^2) m_\pi /M \,,\\
&
\hat e(x,k_\perp^2) = \frac{N_c}{2\pi^3} \left[
\tilde N_{EF}
\frac{\bar{\mathpzc C}_2(\varsigma)}{\varsigma}+
3 N_{EF}\, (1-x)\frac{M^2 \bar{\mathpzc C}_3(\varsigma)}
{\varsigma^2}
\right].
\end{align}
\end{subequations}
This TMD is chiral-odd, \emph{viz}.\ it is associated with an interaction-induced quark chirality flip within the target.  $e(x,k_\perp^2)$ vanishes in the chiral limit, $m_\pi=0$.

\begin{figure}[t]
\includegraphics[clip,width=0.80\linewidth]{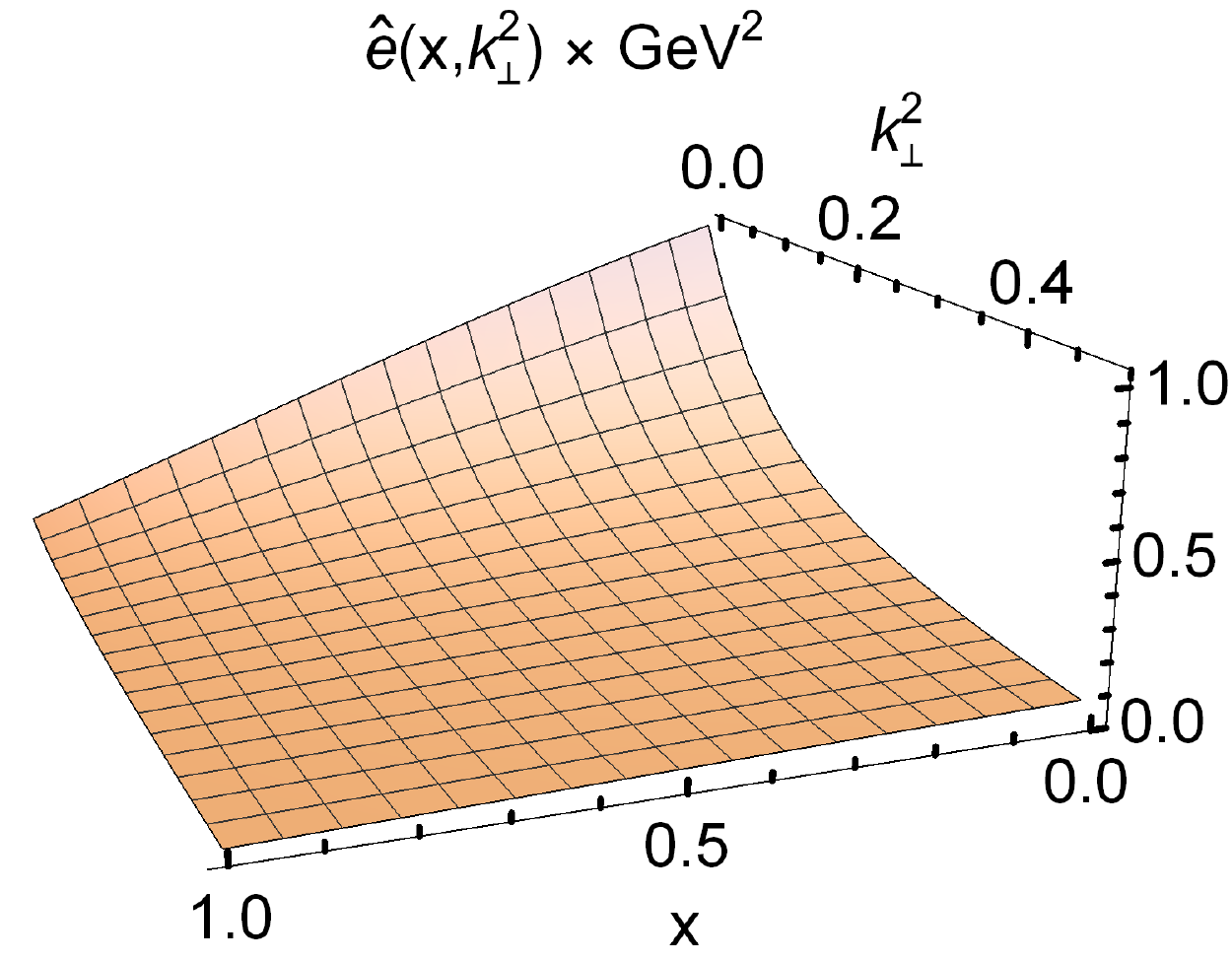}
\includegraphics[clip,width=0.80\linewidth]{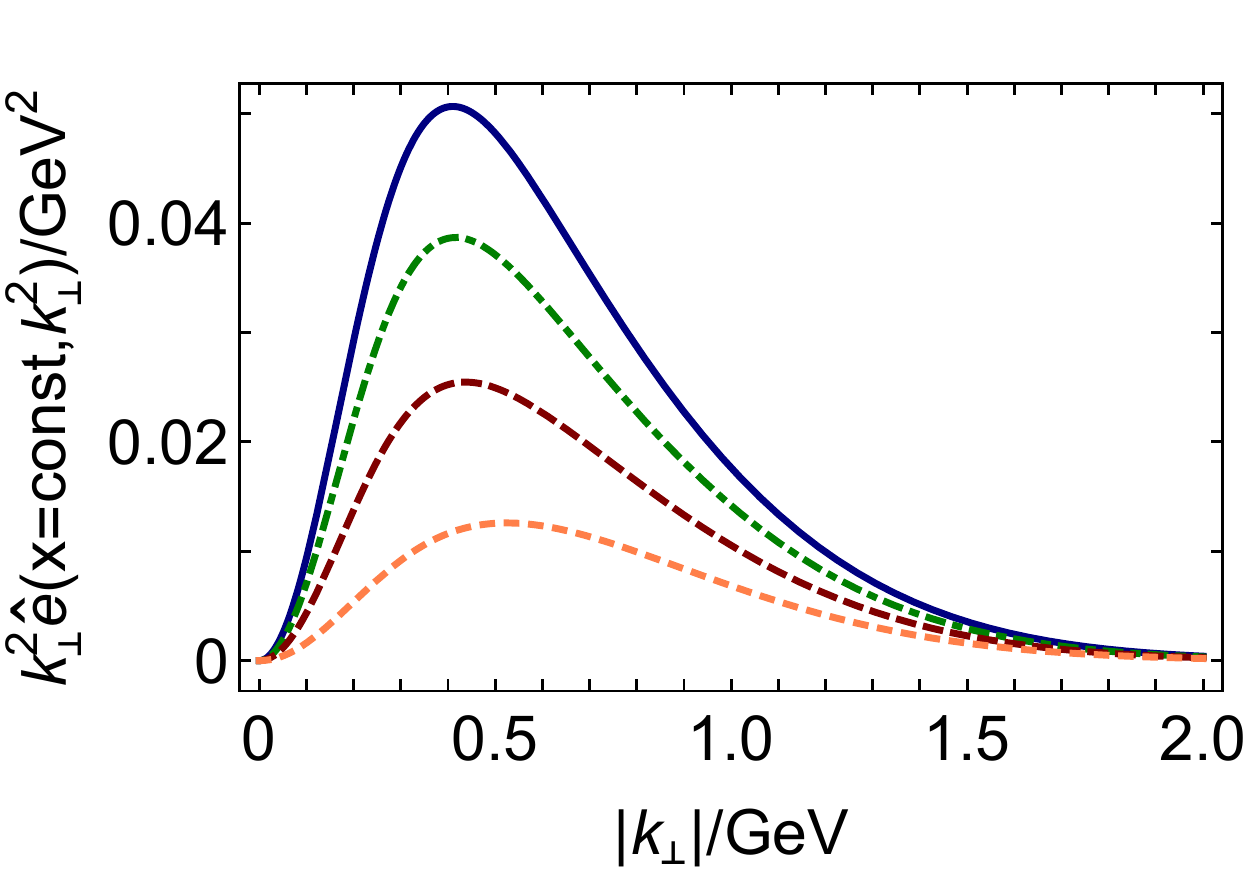}

%
\caption{\label{FigT3TMDe}
\emph{Upper panel}.  Twist-three pion TMD, $\hat e(x,k_\perp^2)$, Eq.\,\eqref{EqpionTMDT3e}.
\emph{Lower panel}.  $k_\perp^2 \hat e(x,k_\perp^2)$ at: $x=0$ -- solid blue curve; $x=1/3$ -- dot-dashed green curve; $x=2/3$ -- dashed red curve; and $x=1$ -- short-dashed orange curve.
Scale is $\zeta_H$, Eq.\,\eqref{HadronScale}.
}
\end{figure}

The upper panel of Fig.\,\ref{FigT3TMDe} depicts the CI result for $\hat e(x,k_\perp^2)$ at the hadronic scale, $\zeta_H$.   The lower panel highlights the $x$-dependence of its $k_\perp^2$ profile:
\begin{equation}
\label{ehatkperp}
\langle k_\perp^2\rangle^{1/2}/{\rm GeV} = 0.385 - 0.109 \,x - 0.0539 \, x^2\,,
\end{equation}
\emph{i.e}.\ the $|k_\perp|$ width ranges from $0.39\,$GeV at $x=0$ to $0.22\,$GeV at $x=1$.   Given the hardness of CI form factors, it is most appropriate to make an internally consistent comparison; hence, we observe that Eq.\,\eqref{ehatkperp} means the width of $e(x,k_\perp^2)$ ranges from 63\% $\to$ 36\% of the width of the chiral-even TMD $f_1(x,k_\perp^2)$, with mean value 51\%.

Comparing the images in Fig.\,\ref{FigT3TMDe} with those in Fig.\,\ref{FigT2TMD}, one sees that $\hat e(x,k_\perp^2)$ is at most two-thirds the size of $f_1(x,k_\perp^2)$ and typically smaller.  In any cross-section, this suppression is compounded by the higher-twist factor $m_\pi/n\cdot P$.

\begin{figure}[t]
\includegraphics[clip,width=0.80\linewidth]{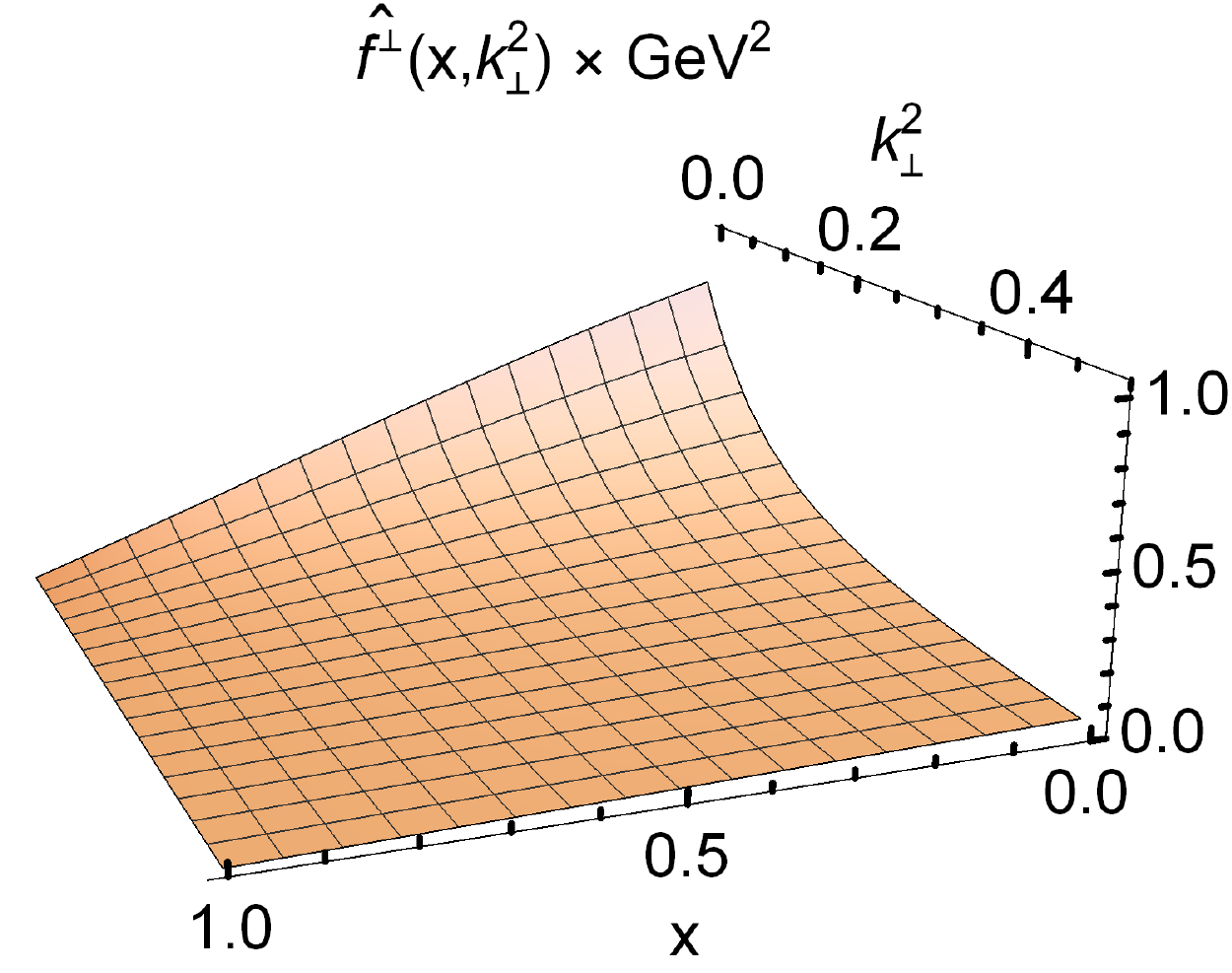}
\includegraphics[clip,width=0.80\linewidth]{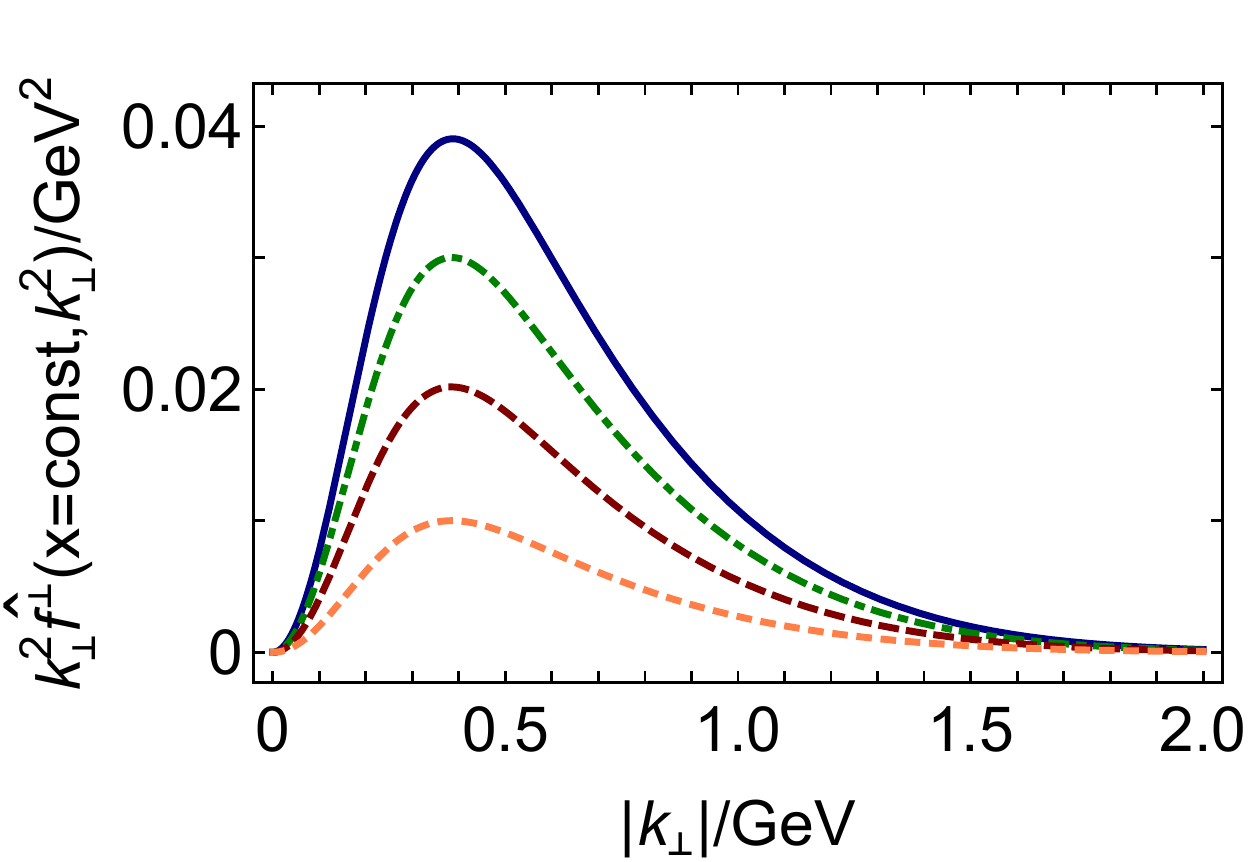}

%
\caption{\label{FigT3TMDf}
\emph{Upper panel}.  Twist-three pion TMD, $\hat f^\perp(x,k_\perp^2)$, Eq.\,\eqref{EqpionTMDT3f}.
\emph{Lower panel}.  $k_\perp^2 \hat f^\perp(x,k_\perp^2)$ at: $x=0$ -- solid blue curve; $x=1/4$ -- dot-dashed green curve; $x=1/2$ -- dashed red curve; and $x=3/4$ -- short-dashed orange curve.
Scale is $\zeta_H$, Eq.\,\eqref{HadronScale}.
}
\end{figure}

The second twist-three TMD, which is chiral-even, may be read from Sec.\,\eqref{AppF2k}:
\begin{subequations}
\label{EqpionTMDT3f}
\begin{align}
f^\perp(x,k_\perp^2) & = F_2^k(x,k_\perp^2,0,0)=: \hat f^\perp(x,k_\perp^2) m_\pi^2 /M^2 , \\
\hat f^\perp(x,k_\perp^2) & =
\frac{3N_c}{2\pi^3} N_{EF} (1-x) \frac{M^2 \bar{\mathpzc C}_3(\varsigma)}{\varsigma^2}\,.
\label{EqpionTMDT3fB}
\end{align}
\end{subequations}
$f^\perp(x,k_\perp^2)$ vanishes in the chiral limit.

$f^\perp(x,k_\perp^2)$ is drawn in Fig.\,\ref{FigT3TMDf}\,--\,upper panel; and the lower panel illustrates the $x$-dependence of its $k_\perp^2$ profile:
\begin{equation}
\label{fhatkperp}
\langle k_\perp^2\rangle^{1/2}/{\rm GeV} = 0.317\,\sqrt{1-x}\,.
\end{equation}
The $|k_\perp|$ width varies from $0.32\,$GeV at $x=0$ to $0$ at $x=1$, owing to the $(1-x)$ factor in Eq.\,\eqref{EqpionTMDT3fB}, \emph{i.e}.\ the width of $f^\perp(x,k_\perp^2)$ ranges from 52\% $\to$ 0\% of the width of the chiral-even TMD $f_1(x,k_\perp^2)$, with mean value 37\%.

Comparison of the images in Fig.\,\ref{FigT3TMDf} with those in Fig.\,\ref{FigT2TMD} reveals that $\hat f^\perp(x,k_\perp^2)$ is not more than two-thirds the size of $f_1(x,k_\perp^2)$ and almost always much smaller.  In any cross-section, this suppression is compounded by the higher-twist factor $(m_\pi/M)(m_\pi/n\cdot P)$.

\subsection{Twist-four TMD}
The CI supports a single twist-four pion TMD, which is chiral-even and can be read from Sec.\,\ref{TMDT4f3}:
\label{EqpionTMDT4f3}
\begin{align}
f_3&(x,k_\perp^2)  = F_3(x,k_\perp^2,0,0)  \nonumber \\
%
& = -\frac{N_c}{2 \pi^3\varsigma} \left[ 2 \tilde N_{EF}\bar{\mathpzc C}_2(\varsigma)
%
+3 N_{EF} (1-x)^2  \frac{m_\pi^2\bar{\mathpzc C}_3(\varsigma)}{\varsigma} \right].
\end{align}
$f_3(x,k_\perp^2)$ is nonzero in the chiral limit so long as the full CI pion Bethe-Salpeter amplitude is used, \emph{i.e}.\ $F_\pi \neq 0$ in Eq.\,\eqref{BSAcontactpion}.

\begin{figure}[t]
\includegraphics[clip,width=0.80\linewidth]{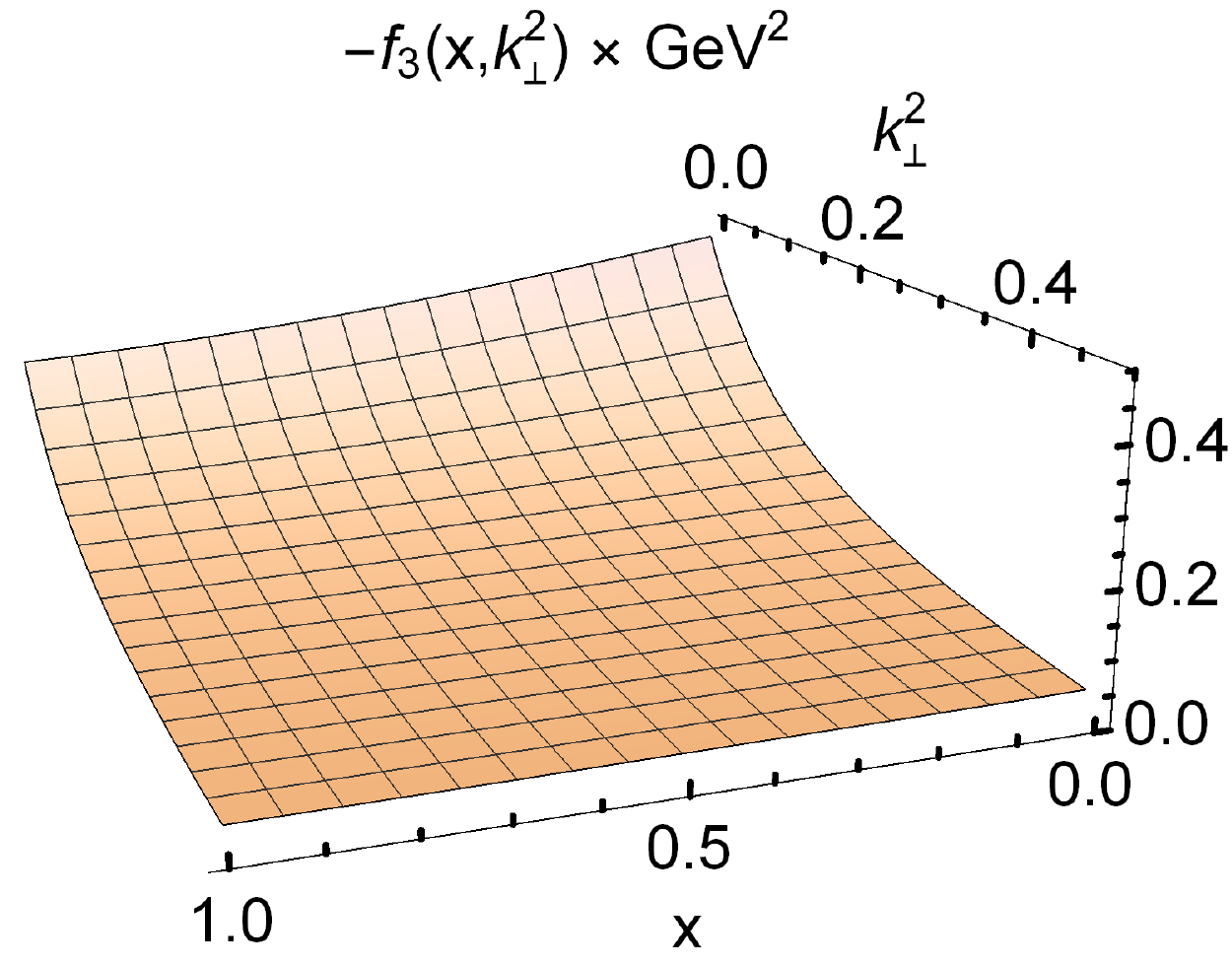}
\includegraphics[clip,width=0.80\linewidth]{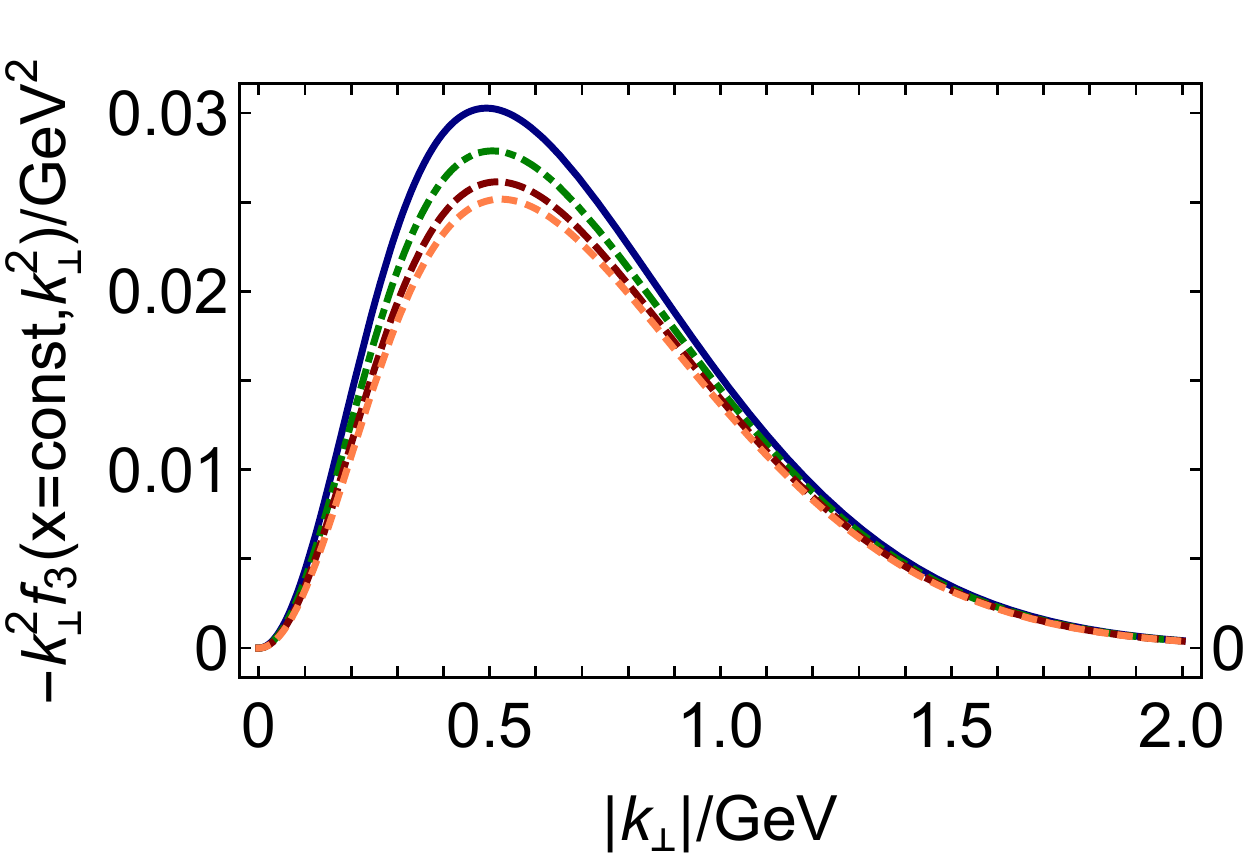}

%
\caption{\label{FigT4TMDf}
\emph{Upper panel}.  Twist-four pion TMD, negative-$f_3(x,k_\perp^2)$, Eq.\,\eqref{EqpionTMDT4f3}.
\emph{Lower panel}.  Negative-$k_\perp^2 f_3(x,k_\perp^2)$ at: $x=0$ -- solid blue curve; $x=1/3$ -- dot-dashed green curve; $x=2/3$ -- dashed red curve; and $x=1$ -- short-dashed orange curve.
Scale is $\zeta_H$, Eq.\,\eqref{HadronScale}.
}
\end{figure}

We depict $\hat f_3(x,k_\perp^2)$ in Fig.\,\ref{FigT4TMDf}\,--\,upper panel; and in the lower panel sketch the $x$-dependence of its $k_\perp^2$ profile:
\begin{equation}
\label{f3hatkperp}
\langle k_\perp^2\rangle^{1/2}/{\rm GeV} = 0.336 - 0.0352 \,x + 0.0129\,x^2.
\end{equation}
Here the $|k_\perp|$ width ranges from $0.34\,$GeV at $x=0$ to $0.31$ at $x=1$, \emph{i.e}.\ the momentum-space breadth of $f_3(x,k_\perp^2)$ ranges from 56\% $\to$ 51\% of the width of $f_1(x,k_\perp^2)$, with mean value 53\%.

Comparing the images in Fig.\,\ref{FigT4TMDf} with those in Fig.\,\ref{FigT2TMD}, it is plain that $\hat f_3(x,k_\perp^2)$ is typically less than one-third the size of $f_1(x,k_\perp^2)$.  This suppression multiplies that introduced into cross-sections by the higher-twist factor $(m_\pi/n\cdot P)^2$.

\section{Wigner Distribution}
\label{SecWigner}
Given that (\emph{a}) GPDs and TMDs can both be obtained directly from Wigner distributions and (\emph{b}) a given Wigner distribution is obtained by computing a Fourier transform of the associated GTMD at $\xi=0$, it is worth presenting a concrete result for the simplest of the Wigner distributions for a dressed-quark in the pion.  To this end, recall Eq.\,\eqref{EqGTMDF1} and consider
\begin{align}
W_{21}(x,k_\perp,b_\perp) & = \int\frac{d^2 \Delta}{(2\pi)^2}
{\rm e}^{ib_\perp\cdot \Delta} \,F_1(x,k_\perp^2,0,-\Delta^2)\,.
\end{align}
Inserting the explicit form of the integrand, one finds
{\allowdisplaybreaks
\begin{align}
& W_{21}(x,k_\perp,b_\perp) \nonumber \\
& =
\frac{N_c  }{4\pi^4} E_\pi[E_\pi - 2 F_\pi]\frac{\bar{\mathpzc C}_2(\varsigma)}{\varsigma} \delta^2(\vec{b}_\perp) \nonumber \\
& \quad +
\frac{N_c }{4\pi^4} E_\pi[E_\pi - 2 F_\pi]\frac{\bar{\mathpzc C}_2(\varsigma)}{\varsigma} \nonumber \\
& \qquad \times
\int_0^\infty \! d\Delta\,\Delta\, J_0(|b_\perp||\Delta|) [P_T -1]
\nonumber \\
& \quad - \frac{3N_c}{8\pi^4} N_{EF}\int_0^\infty d\Delta\,\Delta\, J_0(|b_\perp||\Delta|) P_T
 \nonumber \\
& \qquad \times \int_0^{1-x}\!d\alpha\, [\Delta^2-x(\Delta^2+2 m_\pi^2)] \frac{\bar{\mathpzc C}_3(\varsigma_\alpha)}{\varsigma_\alpha^2}\,, \label{EqWigner}
\end{align}}
\hspace*{-0.5\parindent}where
$\varsigma_\alpha = \varsigma+\alpha(1-x-\alpha)\Delta^2$.  This function has nonzero support on $x \geq 1$.

\begin{figure}[t]
\includegraphics[clip,width=0.80\linewidth]{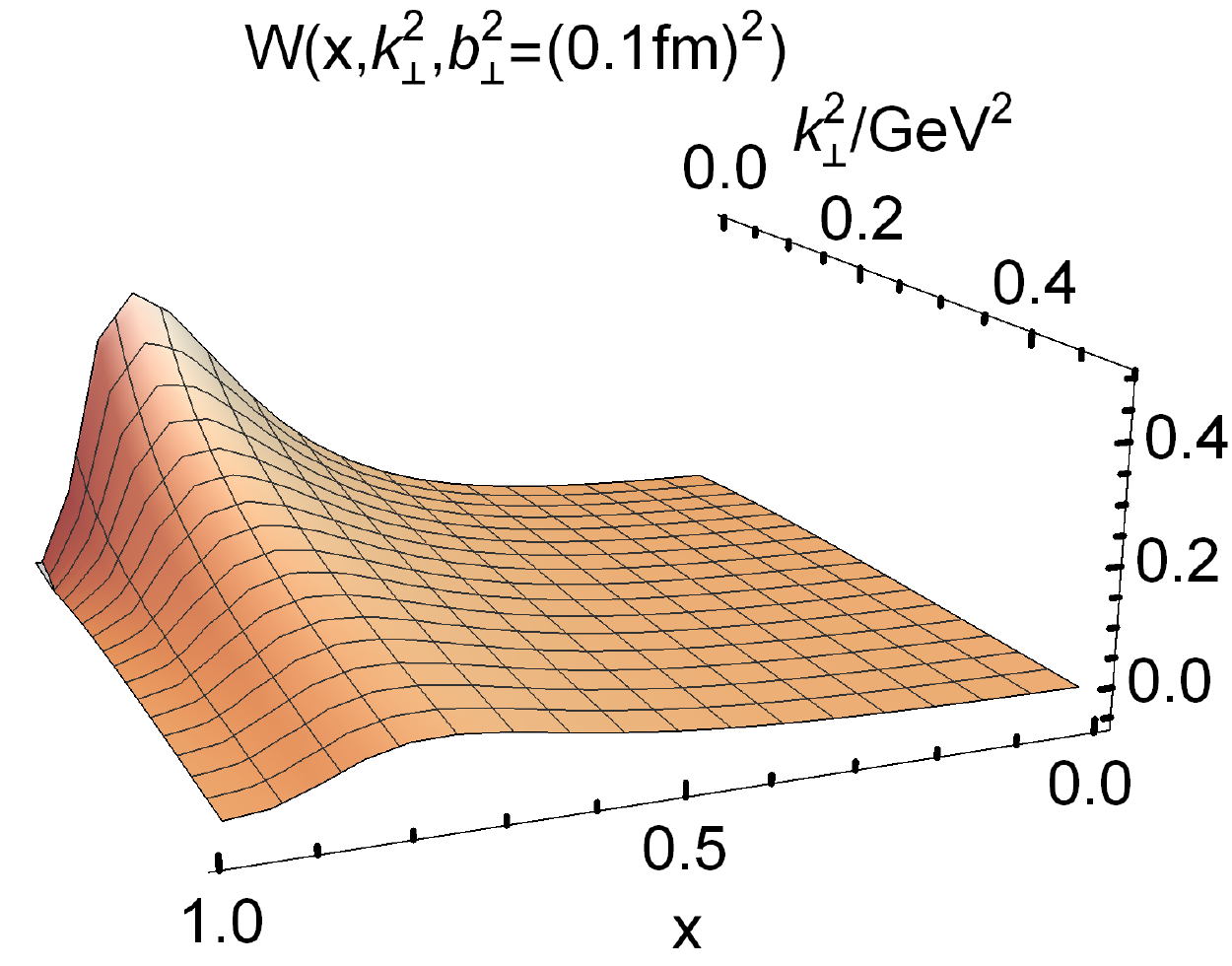}\\[2ex]
\includegraphics[clip,width=0.80\linewidth]{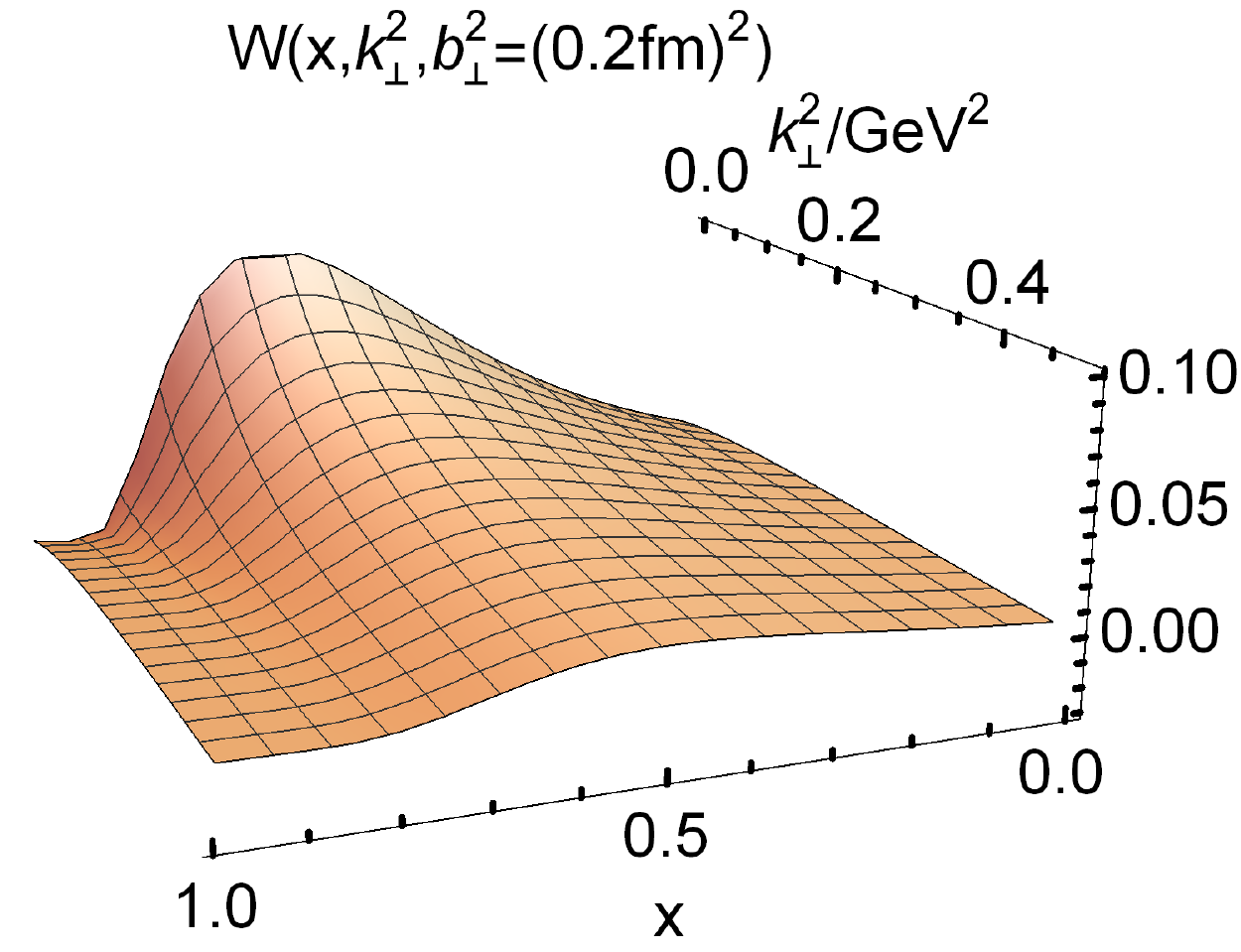}

%
\caption{\label{FigWigner}
Wigner distribution associated with the pion's simplest twist-two dressed-quark GTMD in Eq.\,\eqref{F1start}.  The two panels display different values of $|b_\perp|$, as indicated in the labels.  The $\delta^2(\vec{b}_\perp)$ component -- first line of Eq.\,\eqref{EqWigner} -- is suppressed in the image.
Scale is $\zeta_H$, Eq.\,\eqref{HadronScale}.
}
\end{figure}

The dimensionless Wigner function in Eq.\,\eqref{EqWigner} is plotted in Fig.\,\ref{FigWigner}.  Each panel shows a different value of $|b_\perp|$, \emph{viz}.\ $0.1\,$fm and $0.2\,$fm.  This valence-quark Wigner function is
(\emph{i}) sharply peaked at $(x=1,k_\perp^2=0,b_\perp^2=0)$;
(\emph{ii}) exhibits power-law suppression as $k_\perp^2$ and/or $b_\perp^2$ are increased;
and
(\emph{iii}) is negative on a neighbourhood $(x\simeq 1,k_\perp^2\simeq 0)$.
We anticipate that the analogous Wigner function computed with a realistic interaction will display similar behaviour.

\section{Summary and Perspective}
\label{epilogue}
We used a vector$\,\times\,$vector contact interaction (CI), treated at leading-order in a widely-used symmetry-preserving Dyson-Schwinger equation (DSE) truncation scheme, to calculate an array of twist-two, -three and -four pion GTMDs [Sec.\,\ref{SecT2GTMDs}, Appendices~\ref{AppendixT3}, \ref{AppendixT4}].  Whilst some of the results are particular to the CI, many features are physically relevant, including an observation that the strength and shape of all pion GTMDs are largely set by the scale of emergent hadronic mass (EHM) in the strong interaction.  In a few particular cases for which CI limitations were too conspicuous, we augmented the analysis by appealing to continuum- and lattice-QCD results in order to arrive at realistic illustrations of material points.

Concerning GPDs, we found [Sec.\,\ref{SecVGPDI}] that the pion's $\theta_2$ mass distribution form factor is harder than its electromagnetic form factor, $F_\pi^{\rm em}$; and in turn, $F_\pi$ is harder than the pion's $\theta_1$ gravitational pressure distribution form factor.
Concerning the pressure distribution, the peak value, lying in the neighbourhood of the pion's core, is approximately five-times greater than that in the proton; indeed, it is commensurate with the pressure at the core of a neutron star.  Moreover, the shear pressure achieves its maximum value when the confinement pressure comes to exceed that generated by the forces driving the quark and antiquark away from the core.

The tensor GPD provides information about transversity in the pion; and we found [Sec.\,\ref{SecTensorGPD}] that polarising a pion's dressed quark in the positive-$x$ direction of the light-front-transverse plane produces a clear distortion of the transverse-spin density, shifting its peak in the positive$-y$ direction.  This distortion diminishes as the resolving scale is increased.

The pion's GTMDs also provide direct access to its transverse momentum dependent distribution functions (TMDs); and in the absence of a model for the Wilson line, the CI supports four that are nonzero [Sec.\,\ref{SecTMDs}]: one of twist-two, two twist-three, and one twist-four.  Our calculations indicate that the twist-two TMD, $f_1(x,k_\perp^2)$ is largest in magnitude and possesses the greatest domain of $k_\perp^2$-support.  The twist-three distributions, $e$ and $f^\perp$, are uniformly smaller; and the twist-four TMD, $f_3$, is still smaller.  In any cross-section, these suppressions are compounded by the respective $m_\pi/n\cdot P$ and $(m_\pi/n\cdot P)^2$ twist-expansion factors.

Wigner distributions are a natural complement to GTMDs, providing an intuitive visual aid to expressing and understanding their physical content.  We therefore provided results for a representative example, \emph{viz}.\ that associated with the twist-two GTMD that produces the pion's valence-quark distribution function, and electromagnetic and gravitational form factors [Sec.\ref{SecWigner}].  At the hadronic scale, this Wigner function is sharply peaked in the neighbourhood of $(x=1, k_\perp^2=0, b_\perp^2=0)$ and broadens as the transverse position variable conjugate to the probing momentum transfer, $\vec{b_\perp}$, increases in magnitude.   Similar behaviour should be expected of such Wigner distributions calculated with a realistic interaction.

Several extensions of the work described herein immediately suggest themselves.
(\emph{A}) Kindred analyses for the kaon, which would reveal physical effects on GTMDs that arise from constructive interference between Nature's two mass generating mechanisms: EHM and Higgs-boson induced.
(\emph{B}) Development of a practicable realisation of the Wilson line, because it would enable computation of time-reversal-odd GTMDs, whose comparison with the time-reversal-even functions calculated herein may yield additional insights that could be exploited in studies using realistic interactions.
(\emph{C}) Repeating this analysis using realistic light-front wave functions for the pion (and kaon), whose profiles are known to explain and predict a diverse array of pseudoscalar meson properties.
All these efforts are underway.

\begin{acknowledgements}
%
%
We are grateful for constructive comments and technical assistance from D.~Binosi, C.~Mezrag and J.~Rodr\'{\i}guez-Quintero.
Work supported by:
National Natural Science Foundation of China (under grant nos.\ 11805097, 11175088, 11535005);
Jiangsu Provincial Natural Science Foundation of China (under grant no. BK20180323);
and Jiangsu Province \emph{Hundred Talents Plan for Professionals}.
\end{acknowledgements}

\appendix

\section{Useful Formulae}
\label{AppFormulae}
Eq.\,\eqref{gapactual} is the first of many integrals appearing herein whose regularised values are expressed in terms of incomplete gamma-functions.  In general ($n\in \mathbb Z$, $n\geq 0$):
\begin{subequations}
\label{Cndef}
\begin{align}
\label{C0def}
{\cal C}_0(\sigma) &=
\sigma \big[\Gamma(-1,\sigma \tau_{\rm uv}^2) - \Gamma(-1,\sigma \tau_{\rm ir}^2)\big]\,,\\
{\cal C}_n(\sigma) & = (-)^n \frac{\sigma^n}{n!}\frac{d^n}{d\sigma^n} {\cal C}_0(\sigma)\,, \\
\overline{\cal C}_n(\sigma) & = \frac{1}{\sigma} {\cal C}_n(\sigma)\,.
\end{align}
\end{subequations}
They can usefully be illustrated with simple examples:
\begin{subequations}
\label{C012def}
\begin{align}
\overline{\cal C}_0(\sigma) &
= \Gamma(-1,\sigma \tau_{\rm ir}^2) - \Gamma(-1,\sigma \tau_{\rm uv}^2)\,,\\
\overline{\cal C}_1(\sigma) & = \Gamma(0,\sigma \tau_{\rm ir}^2) - \Gamma(0,\sigma \tau_{\rm uv}^2)\,,\\
2\, \overline{\cal C}_2(\sigma) & = \sigma \frac{d^2}{d\sigma^2} {\cal C}_0(\sigma)
= \Gamma(1,\sigma \tau_{\rm ir}^2) - \Gamma(1,\sigma \tau_{\rm uv}^2)\,.
\end{align}
\end{subequations}
In general,
\begin{equation}
n! \, \overline{\cal C}_n(\sigma) =
\Gamma(n-1,\sigma \tau_{\rm ir}^2) - \Gamma(n-1,\sigma \tau_{\rm uv}^2)\,.
\end{equation}

Such expressions are useful, \emph{e.g}.\ in expressing the Bethe-Salpeter kernel in Eq.\,\eqref{bsefinalE}:
{\allowdisplaybreaks
\begin{subequations}
\label{fgKernel}
\begin{align}
\nonumber
&{\cal K}_{EE}^{\pi}=
\int_0^1d\alpha \bigg\{
{\cal C}_0(\omega( \alpha, Q^2))  \\
&+ \bigg[ M^2 -\alpha \bar\alpha Q^2 - \omega( \alpha, Q^2)\bigg]\overline{\cal C}_1(\omega(\alpha, Q^2))\bigg\}\, ,\\
&{\cal K}_{EF}^{\pi} =Q^2 \int_0^1d\alpha\, \overline{\cal C}_1(\omega(\alpha, Q^2)),\\
&{\cal K}_{FE}^{\pi} = \tfrac{1}{2} M^2\int_0^1d\alpha\, \overline{\cal C}_1(\omega(\alpha, Q^2))\, ,\\
&{\cal K}_{FF}^{\pi} = - M^2 \int_0^1d\alpha\,  \overline{\cal C}_1(\omega(\alpha, Q^2))\,.
\end{align}
\end{subequations}}

We recall here that Eq.\,\eqref{bsefinalE} is an eigenvalue problem with a solution for $Q^2=-m_{\pi}^2$, at which point the eigenvector is the pion's Bethe-Salpeter amplitude.  When computing observables, one must employ the canonically normalised amplitude, \emph{viz}.\ $\Gamma_\pi$ rescaled such that
\begin{equation}
\label{normcan}
1=\left. \frac{d}{d Q^2}\Pi_{\pi}(Z,Q)\right|_{Z=Q},
\end{equation}
where
\begin{equation}
\Pi_{\pi}(Z,Q)= 6 {\rm tr}_{\rm D} \!\! \int\! \frac{d^4\ell}{(2\pi)^4} \Gamma_{\pi}(-Z)
 S_f(\ell+Q) \, \Gamma_{\pi}(Z)\, S_g(\ell)\,. \label{normcan2}
\end{equation}
In the chiral limit, \emph{viz}.\ using solutions obtained with $m=0$ in Eq.\,\eqref{GapEqn}, Eqs.\,\eqref{normcan}, \eqref{normcan2} impose \cite{GutierrezGuerrero:2010md}:
\begin{equation}
1 = \frac{3}{4\pi^2}\frac{1}{M^2}{\cal C}_1(M^2) E_\pi [E_\pi - 2 F_\pi]\,.
\end{equation}

The function $\omega(\alpha,Q^2)$ is defined in Eq.\,\eqref{eq:omega}.  Similar arguments appear in the expressions for various pion GTMDs.  We list them here.
{\allowdisplaybreaks
\begin{subequations}
\label{Eqsigmas}
\begin{align}
\sigma_1^{z,u} & = z + M^2 -\frac{(x+ u\xi)(1-x)}{(1+u\xi)^2} m_\pi^2 , \\
\sigma_2^z & = z + M^2 -\frac{1}{4}\left(1+\frac{x}{\xi}\right)\left(1-\frac{x}{\xi}\right)t \,,\\
\sigma_3^z & = z + M^2 -\alpha \bar\alpha m_\pi^2 \nonumber\\
& - [\xi+x-\alpha(1+\xi)][\xi-x+\alpha(1-\xi)]\frac{t}{4\xi^2}\,,\\
\sigma_4^z & = \sigma_1^{z,0} - \alpha (1-\alpha-x) t\,,\\
\sigma_5 & = M^2 - x (1-x) t\,,\\
\sigma_6 & = M^2 - (x+y)(1-x-y) m_\pi^2 - x y t \,.
\end{align}
When describing TMDs, we also use
\begin{equation}
\varsigma := \sigma_1^{k_\perp^2,0} = k_\perp^2 + M^2 - x(1-x)m_\pi^2 .
\end{equation}
\end{subequations}
}

\section{Twist Three GTMDs}
\label{AppendixT3}
Here we gather CI results for the pion's dressed-quark twist-three GTMDs, of which there are six, generated by the following matrix insertions in Eq.\,\eqref{GTMD}:
\begin{align}
{\mathpzc H} &
\to \{ {\mathpzc H}_1 = 1 \,,\,
{\mathpzc H}_2 = i\gamma_5\,,\,
{\mathpzc H}_3 = i\gamma_j\,,\,
{\mathpzc H}_4 = i\gamma_j\gamma_5 \,,\nonumber \\
& \qquad
{\mathpzc H}_5 = i\gamma_5 \sigma_{ij} \,,\,
{\mathpzc H}_6 = i \gamma_5 \sigma_{\mu\nu} n_\mu \bar{n}_\nu \}.
\end{align}
Specifically,
mapping into Euclidean metric, suppressing the argument, $(P,x,\vec{k}_\perp,\Delta,N;\eta)$, of each GTMD on the left-hand-side, and writing $\check k = k/M$, $\check \Delta = \Delta/M$:
{\allowdisplaybreaks
\begin{subequations}
\begin{align}
W^{[{\mathpzc H}_1]} &\to \frac{M}{n\cdot P} \, E_2(x,k_\perp^2,\xi,t) \,, \\
W^{[{\mathpzc H}_2]} & \to \frac{M}{n\cdot P} \, i\varepsilon_{ij}^{\perp} \check k_i \check\Delta_j \tilde{E_2} (x,k_\perp^2,\xi,t) \,, \\
W^{[{\mathpzc H}_3]} &\to \frac{M}{n\cdot P} [ \check k_iF_2^k(x,k_\perp^2,\xi,t) \nonumber \\
& \qquad\qquad + \check \Delta_i F_2^{\Delta} (x,k_\perp^2,\xi,t)]\,, \\
W^{[{\mathpzc H}_4]} &\to \frac{M}{n\cdot P} [ i\varepsilon_{ij}^ {\perp} \check k_i G_2^k(x,k_\perp^2,\xi,t) \nonumber \\
& \qquad\qquad +  i\varepsilon_{ij}^{\perp}\check \Delta_i G_2^{\Delta}(x,k_\perp^2,\xi,t)]\,,\\
W^{[{\mathpzc H}_5]}&\to\frac{M}{n\cdot P}\, i\varepsilon_{ij}^{\perp}H_2 (x,k_\perp^2,\xi,t)\,, \\
W^{[{\mathpzc H}_6]} &\to \frac{M}{n\cdot P} \, i\varepsilon_{ij}^{\perp} \check k_i \check \Delta_j\tilde{H_2} (x,k_\perp^2,\xi,t)\,.
\end{align}
\end{subequations}

\subsection{$\mathbf{E_2}$}
\unskip
\label{AppE2}
\begin{align}
E_2(x,& k_\perp^2,\xi,t)  =  \nonumber \\
& \times \bar P_{\rm T} \left[E_\pi^2 \,F^{EE} +  E_\pi F_\pi \, F^{EF}  + F_\pi^2 \,F^{FF}\right] \,,
\label{EqGTMDE2}
\end{align}
where $\bar P_{\rm T}= [\theta_{\bar\xi\xi} + P_{\rm T}(-t)(1-\theta_{\bar\xi\xi}) ] $ and $(r=k_\perp^2)$:
{\allowdisplaybreaks
\begin{subequations}
\begin{align}
 F^{EE}(x,r,\xi,t)  &=T_{12}^{EE}+\frac{N_c}{4\pi^3} \frac{1}{\sigma_2^r}\bar {\cal C}_2(\sigma_2^{r})\frac{\theta_{\bar\xi\xi} }{\xi} \nonumber \\
 F^{EF}(x,r,\xi,t)  & = T_{11}^{EF} - 4 T_{12}^{EE}\nonumber \\
 &-\frac{N_c}{4\pi^3} \frac{1}{\sigma_2^r}\bar {\cal C}_2(\sigma_2^{r})\frac{\theta_{\bar\xi\xi} t}{M^2 \xi} \,,\\
 F^{FF}(x,r,\xi,t)  & =-2\, T_{11}^{EF}+4 T_{12}^{EE} \nonumber \\
 & \quad + \frac{N_c}{4\pi^3} \frac{1}{\sigma_2^r}\bar {\cal C}_2(\sigma_2^{r}) \frac{\theta_{\bar\xi\xi}t}{M^2\xi} \nonumber \\
 &\quad +\frac{N_c}{8\pi^3} \frac{1}{\sigma_2^r}\bar {\cal C}_2(\sigma_2^{r}) \frac{\theta_{\bar\xi\xi}x t}{M^2 \xi^3} \,,
\end{align}
\end{subequations}}
with
{\allowdisplaybreaks
\begin{subequations}
\begin{align}
  T_{11}^{EF}& (x,r,\xi,t) = \nonumber \\
 & \quad \frac{N_c}{4\pi^3}\frac{m_{\pi}^2 }{ M^2(1+\xi)}
 \frac{\theta_{\bar \xi 1}}{\sigma_1^{r,1}}\bar {\cal C}_2(\sigma_1^{r,1})\nonumber\\
 &+  \frac{N_c}{4\pi^3}\frac{m_{\pi}^2 }{M^2(1-\xi)}\frac{\theta_{\xi 1}}{\sigma_1^{r,-1}}\bar {\cal C}_2(\sigma_1^{r,-1}) \,, \\
  T_{12}^{EE}&(x,r,\xi,t) = \nonumber \\
 & \quad \frac{3 N_c}{8\pi^3} \frac{\left(2m_{\pi}^2-t\right)}{\xi} \int_0^1 d\alpha\,
\frac{\theta_{\alpha\xi}}{[\sigma_3^r]^2}\bar {\cal C}_3(\sigma_3^{r}) \,.
\end{align}
\end{subequations}}

\subsection{$\mathbf{\tilde E_2}$}
\unskip
\begin{align}
\tilde{E}_2(x,& k_\perp^2,\xi,t)  = 0\,. \label{EqGTMDEt2}
\end{align}

\subsection{$\mathbf{F_2^k}$}
\unskip
\label{AppF2k}
\begin{align}
F_2^k (x,& k_\perp^2,\xi,t)  =  \nonumber \\
& \times \bar P_{\rm T} \left[E_\pi^2 \,F^{EE} +  E_\pi F_\pi \, F^{EF}  + F_\pi^2 \,F^{FF}\right] \,,
\label{EqGTMDF2l}
\end{align}
where 
{\allowdisplaybreaks
\begin{subequations}
\begin{align}
 F^{EE}(x,r,\xi,t)  &=  T_{31}^{EE} + \frac{N_c}{4\pi^3} \frac{1}{\sigma_2^r}\bar {\cal C}_2(\sigma_2^{r})\frac{\theta_{\bar\xi\xi} }{\xi}\,, \\
 F^{EF}(x,r,\xi,t)  & =  - 4 T_{31}^{EE} \,,\\
 F^{FF}(x,r,\xi,t)  & =4 T_{31}^{EE} +\frac{N_c}{8\pi^3} \frac{1}{\sigma_2^r}\bar {\cal C}_2(\sigma_2^{r}) \frac{\theta_{\bar\xi\xi}x t}{M^2\xi^3} \,,
\end{align}
\end{subequations}
with
\begin{align}
  T_{31}^{EE}&(x,r,\xi,t) = \nonumber \\
 &   \frac{3 N_c}{8\pi^3} \frac{(2m_{\pi}^2-t)}{\xi} \int_0^1 d\alpha\,
\frac{\theta_{\alpha\xi}}{[\sigma_3^r]^2}\bar {\cal C}_3(\sigma_3^{r}) \,.
\end{align}}

\subsection{$\mathbf{F_2^\Delta}$}
\unskip
\begin{align}
 F_2^{\Delta} (x,& k_\perp^2,\xi,t)  =  \nonumber \\
& \times \bar P_{\rm T} \left[E_\pi^2 \,F^{EE} +  E_\pi F_\pi \, F^{EF}  + F_\pi^2 \,F^{FF}\right] \,,
\label{EqGTMDFq}
\end{align}
where 
{\allowdisplaybreaks
\begin{subequations}
\begin{align}
 F^{EE}(x,r,\xi,t)  &= T_{41}^{EE} + T_{42}^{EE} \nonumber \\
&  \quad -\frac{N_c}{8\pi^3} \frac{1}{\sigma_2^r}\bar {\cal C}_2(\sigma_2^{r})\frac{\theta_{\bar\xi\xi} x}{\xi^2}\,, \\
 F^{EF}(x,r,\xi,t)  & = - 2 \, T_{41}^{EE} - 4 T_{42}^{EE} \,,\\
 F^{FF}(x,r,\xi,t)  & =4 T_{42}^{EE} + \frac{N_c}{16\pi^3} \frac{1}{\sigma_2^r}\bar {\cal C}_2(\sigma_2^{r}) \nonumber \\
 & \quad \times \frac{\theta_{\bar\xi\xi}}{M^2\xi^2} \left[\frac{1}{2}t-\frac{3x^2}{ 2\xi^2 }t-2M^2\right] ,
\end{align}
\end{subequations}}
with
{\allowdisplaybreaks
\begin{subequations}
\begin{align}
  T_{41}^{EE}& (x,r,\xi,t) =
  -\frac{N_c}{8\pi^3} \left[
 \frac{1}{(1+\xi)}\frac{\theta_{\bar \xi 1}}{\sigma_1^{r,1}}\bar {\cal C}_2(\sigma_1^{r,1}) \right. \nonumber \\
 & \left.
 \quad \quad \quad \quad - \frac{1}{(1-\xi)}\frac{\theta_{\xi 1}}{\sigma_1^{r,-1}}\bar {\cal C}_2(\sigma_1^{r,-1}) \right] \,, \\
  T_{42}^{EE}&(x,r,\xi,t) = -\frac{3 N_c}{16\pi^3}  \nonumber \\
 &   \times  \int_0^1 d\alpha \left[\frac{x-\alpha}{\xi^2}(2m_{\pi}^2-t) \right]\,
\frac{\theta_{\alpha\xi}}{[\sigma_3^r]^2}\bar {\cal C}_3(\sigma_3^{r}) \,.
\end{align}
\end{subequations}}

\subsection{$\mathbf{G_2^k}$}
\unskip
\begin{align}
G_2^k(x,& k_\perp^2,\xi,t)  =  \nonumber \\
& \times \bar P_{\rm T} \left[E_\pi^2 \,F^{EE} +  E_\pi F_\pi \, F^{EF}  + F_\pi^2 \,F^{FF}\right] \,,
\label{EqGTMDGl}
\end{align}
where 
\begin{subequations}
\begin{align}
 F^{EE}(x,r,\xi,t)  &= T_{51}^{EE} \,, \\
 F^{EF}(x,r,\xi,t)  & =- 4 T_{51}^{EE} \,,\\
 F^{FF}(x,r,\xi,t)  & =4 T_{51}^{EE} \nonumber \\
 & \quad + \frac{N_c}{4\pi^3} \frac{1}{\sigma_2^r}\bar {\cal C}_2(\sigma_2^{r}) \frac{\theta_{\bar\xi\xi}(4m_{\pi}^2-t)}{M^2} \,,
\end{align}
\end{subequations}
with
\begin{align}
  T_{51}^{EE}&(x,r,\xi,t) = \nonumber \\
 &  - \frac{3 N_c}{4\pi^3} (4m_{\pi}^2-t) \int_0^1 d\alpha\,
\frac{\theta_{\alpha\xi}}{[\sigma_3^r]^2}\bar {\cal C}_3(\sigma_3^{r}) \,.
\end{align}

\subsection{$\mathbf{G_2^\Delta}$}
\unskip
\begin{align}
G_2^{\Delta}(x,& k_\perp^2,\xi,t)  =  \nonumber \\
& \times \bar P_{\rm T} \left[E_\pi^2 \,F^{EE} +  E_\pi F_\pi \, F^{EF}  + F_\pi^2 \,F^{FF}\right] \,,
\label{EqGTMDGq}
\end{align}
where
\begin{subequations}
\begin{align}
 F^{EE}(x,r,\xi,t)  &= T_{61}^{EE}\,, \\
 F^{EF}(x,r,\xi,t)  & =- 4 T_{61}^{EE} \,,\\
 F^{FF}(x,r,\xi,t)  & =4 T_{61}^{EE} +\frac{N_c}{8\pi^3} \frac{1}{\sigma_2^r}\bar {\cal C}_2(\sigma_2^{r}) \frac{\theta_{\bar\xi\xi}xt}{M^2\xi^3}\,,
\end{align}
\end{subequations}
with
\begin{align}
  T_{61}^{EE}&(x,r,\xi,t) =  \nonumber \\
 &  - \frac{3 N_c}{8\pi^3}\int_0^1 d\alpha\,
\frac{ t(x-\alpha)}{\xi^3} \frac{\theta_{\alpha\xi}}{[\sigma_3^r]^2}\bar {\cal C}_3(\sigma_3^{r}) \,.
\end{align}

\subsection{$\mathbf{H_2}$}
\unskip
\begin{align}
 H_2(x,& k_\perp^2,\xi,t)  =  \nonumber \\
& \times \bar P_{\rm T} \left[E_\pi^2 \,F^{EE} +  E_\pi F_\pi \, F^{EF}  + F_\pi^2 \,F^{FF}\right] \,,
\label{EqGTMDH2}
\end{align}
where 
\begin{subequations}
\begin{align}
 F^{EE}(x,r,\xi,t)  &=  T_{72}^{EE}\,, \\
 F^{EF}(x,r,\xi,t)  & =  T_{71}^{EF} - 4 T_{72}^{EE}\nonumber\\
 &\quad- \frac{N_c}{2\pi^3} \frac{1}{\sigma_2^r}\bar {\cal C}_2(\sigma_2^{r}) \frac{\theta_{\bar\xi\xi}xt}{M^2\xi^2}  \,,\\
 F^{FF}(x,r,\xi,t)  & = -2T_{71}^{EF}+4 T_{72}^{EE} \nonumber \\
 &\quad + \frac{N_c}{2\pi^3} \frac{1}{\sigma_2^r}\bar {\cal C}_2(\sigma_2^{r}) \frac{\theta_{\bar\xi\xi}xt}{M^2\xi^2} \nonumber\\
 & \quad - \frac{N_c}{4\pi^3} \frac{1}{\sigma_2^r}\bar {\cal C}_2(\sigma_2^{r}) \frac{\theta_{\bar\xi\xi}(4m_{\pi}^2-t)}{M^2}  \,,
\end{align}
\end{subequations}
with
{\allowdisplaybreaks\begin{subequations}
\begin{align}
  T_{71}^{EF}& (x,r,\xi,t) = \nonumber \\
 &
 -\frac{N_c}{2\pi^3}\frac{(1-2x-\xi)m_{\pi}^2}{M^2(1+\xi)^2 } \frac{\theta_{\bar \xi 1}}{\sigma_1^{r,1}}\bar {\cal C}_2(\sigma_1^{r,1})\nonumber \\
 &+   \frac{N_c}{2\pi^3}\frac{(1-2x+\xi)m_{\pi}^2}{M^2(1-\xi)^2} \frac{\theta_{\xi 1}}{\sigma_1^{r,-1}}\bar {\cal C}_2(\sigma_1^{r,-1})  \,, \\
  T_{72}^{EE}&(x,r,\xi,t) = \nonumber \\
 &   \frac{3 N_c}{4\pi^3} (4m_{\pi}^2-t)  \int_0^1 d\alpha\,
\frac{\theta_{\alpha\xi}}{[\sigma_3^r]^2}\bar {\cal C}_3(\sigma_3^{r}) \,.
\end{align}
\end{subequations}}

\subsection{$\mathbf{\tilde H_2}$}
\unskip
\begin{align}
\tilde{H}_2(x,& k_\perp^2,\xi,t)  =  \bar P_{\rm T} \left[  E_\pi F_\pi \, F^{EF}  + F_\pi^2 \,F^{FF}\right] \,,
\end{align}
where 
\begin{subequations}
\begin{align}
 F^{EF}(x,r,\xi,t)  & = T_{81}^{EF} -\frac{N_c}{2\pi^3} \frac{1}{\sigma_2^r}\bar {\cal C}_2(\sigma_2^{r})\frac{\theta_{\bar\xi\xi} }{\xi}\,, \\
 F^{FF}(x,r,\xi,t)  & =-2T_{81}^{EF} +\frac{N_c}{2\pi^3} \frac{1}{\sigma_2^r}\bar {\cal C}_2(\sigma_2^{r})\frac{\theta_{\bar\xi\xi}}{\xi}\,,
\end{align}
\end{subequations}
with
\begin{align}
  T_{81}^{EF}&(x,r,\xi,t) = \frac{N_c}{4\pi^3}\frac{1}{ (1+\xi)}\frac{\theta_{\bar \xi 1}}{\sigma_1^{r,1}}\bar {\cal C}_2(\sigma_1^{r,1})\nonumber \\
 & +    \frac{N_c}{4\pi^3}\frac{1}{ (1-\xi)}\frac{\theta_{\xi 1}}{\sigma_1^{r,-1}}\bar {\cal C}_2(\sigma_1^{r,-1})\,.
\end{align} 

\section{Twist Four GTMDs}
\label{AppendixT4}
Here we list the CI results for the pion's dressed-quark twist-four GTMDs, of which there are four, generated by the following matrix insertions in Eq.\,\eqref{GTMD}:
\begin{align}
{\mathpzc H} \to \{ {\mathpzc H}_1 = i\gamma\cdot \bar{n} \,,\,
{\mathpzc H}_2 = i\gamma\cdot \bar{n} \gamma_5\,,\,
{\mathpzc H}_3 = i\gamma_5 \sigma_{j \mu}\bar{n}_\mu  \}.
\end{align}
Mapping into Euclidean metric:
\begin{subequations}
\begin{align}
W^{[{\mathpzc H}_1]} & \to \frac{M^2}{(P\cdot n)^2} \,  F_3(x,k_\perp^2,\xi,t) \,, \\
W^{[{\mathpzc H}_2]} &\to \frac{M^2}{(P\cdot n)^2} \,  i\varepsilon_{ij}^{\perp} \check k_i \check \Delta_j\tilde{G_3}(x,k_\perp^2,\xi,t) \,, \\
W^{[{\mathpzc H}_3]} &\to\frac{M^2}{(P\cdot n)^2} \,  [i\varepsilon_{ij}^{\perp} \check k_iH_3^k (x,k_\perp^2,\xi,t) \,, \\
& \qquad\qquad\quad + i\varepsilon_{ij}^{\perp}\check \Delta_i H_3^{\Delta} (x,k_\perp^2,\xi,t)]\,.
\end{align}
\end{subequations}

\subsection{$\mathbf{F_3}$}
\unskip
\label{TMDT4f3}
\begin{align}
F_3(x,& k_\perp^2,\xi,t)  =  \nonumber \\
& \times \bar P_{\rm T} \left[E_\pi^2 \,F^{EE} +  E_\pi F_\pi \, F^{EF}  + F_\pi^2 \,F^{FF}\right] \,,
\label{EqGTMDF3}
\end{align}
where
\begin{subequations}
\begin{align}
 F^{EE}&(x,r,\xi,t)  = \tilde{T}_{11}^{EE} + \tilde{T}_{12}^{EE} \nonumber \\
&  \quad + \frac{N_c}{16\pi^3} \frac{1}{\sigma_2^r}\bar {\cal C}_2(\sigma_2^{r})\frac{\theta_{\bar\xi\xi} x(4m_{\pi}^2-t)}{M^2\xi}\nonumber \\
&  \quad + \frac{N_c}{8\pi^3} \frac{1}{\sigma_2^r}\bar {\cal C}_2(\sigma_2^{r})\frac{\theta_{\bar\xi\xi} xt}{M^2\xi^3}\,, \\
 F^{EF}&(x,r,\xi,t)  = - 2 \, \tilde{T}_{11}^{EE} - 4 \tilde{T}_{12}^{EE} \,,\\
 F^{FF}&(x,r,\xi,t)  =4 \tilde{T}_2^{EE} \nonumber \\
 & \quad + \frac{N_c}{32\pi^3} \frac{1}{\sigma_2^r}\bar {\cal C}_2(\sigma_2^{r}) \frac{\theta_{\bar\xi\xi}t(4m_{\pi}^2-t)}{M^4\xi}  \left[1-\frac{x^2}{\xi^2} \right]\,,
\end{align}
\end{subequations}
with
{\allowdisplaybreaks
\begin{subequations}
\begin{align}
  \tilde{T}_{11}^{EE}& (x,r,\xi,t) = \nonumber \\
 &  -\frac{N_c}{16\pi^3}
 \frac{(1-\xi)(4m_{\pi}^2-t) }{M^2(1+\xi)}\frac{\theta_{\bar \xi 1}}{\sigma_1^{r,1}}\bar {\cal C}_2(\sigma_1^{r,1})\nonumber\\
 &-  \frac{N_c}{16\pi^3}\frac{(4m_{\pi}^2-t) (1+\xi)}{M^2(1-\xi)}  \frac{\theta_{\xi 1}}{\sigma_1^{r,-1}}\bar {\cal C}_2(\sigma_1^{r,-1})  \,, \\
  \tilde{T}_{12}^{EE}&(x,r,\xi,t) = \nonumber \\
 & \quad   \frac{3 N_c}{32\pi^3} \int_0^1 d\alpha \frac{((2\alpha -x)(t-2m_{\pi}^2)- t)(4m_{\pi}^2-t)}{M^2\xi}\,\nonumber\\
&\times \frac{\theta_{\alpha\xi}}{[\sigma_3^r]^2}\bar {\cal C}_3(\sigma_3^{r}) \nonumber\\
 &+\frac{3 N_c}{16\pi^3} \int_0^1 d\alpha \frac{(2m_{\pi}^2-t) (x-\alpha)t}{M^2\xi^3}\,
\frac{\theta_{\alpha\xi}}{[\sigma_3^r]^2}\bar {\cal C}_3(\sigma_3^{r}) \,.
\end{align}
\end{subequations}}

\subsection{$\mathbf{\tilde{G}_3}$}
\unskip
\begin{align}
\tilde{G}_3(x,& k_\perp^2,\xi,t)  =  \nonumber \\
& \times \bar P_{\rm T} \left[E_\pi^2 \,F^{EE} +  E_\pi F_\pi \, F^{EF}  + F_\pi^2 \,F^{FF}\right] \,,
\label{EqGTMDG3}
\end{align}
where
{\allowdisplaybreaks
\begin{subequations}
\begin{align}
 F^{EE}(x,r,\xi,t)  &=\tilde{T}_{21}^{EE}\,, \\
 F^{EF}(x,r,\xi,t)  & = - 4 \tilde{T}_{21}^{EE} \,,\\
 F^{FF}(x,r,\xi,t)  & =4 \tilde{T}_{21}^{EE}- \frac{N_c}{16\pi^3} \frac{1}{\sigma_2^r}\bar {\cal C}_2(\sigma_2^{r}) \frac{\theta_{\bar\xi\xi}}{M^2 \xi}\,,
\end{align}
\end{subequations}}
with
\begin{align}
  \tilde{T}_{21}^{EE}&(x,r,\xi,t) = -\frac{3 N_c}{16\pi^3}\frac{1}{\xi} \int_0^1 d\alpha\,
\frac{\theta_{\alpha\xi}}{[\sigma_3^r]^2}\bar {\cal C}_3(\sigma_3^{r}) \,.
\end{align}

\subsection{$\mathbf{H_3^\Delta}$}
\unskip
\begin{align}
H_3^{\Delta}(x,& k_\perp^2,\xi,t)  =  \nonumber \\
& \times \bar P_{\rm T} \left[E_\pi^2 \,F^{EE} +  E_\pi F_\pi \, F^{EF}  + F_\pi^2 \,F^{FF}\right] \,,
\label{EqGTMDH3}
\end{align}
where
\begin{subequations}
\begin{align}
 F^{EE}(x,r,\xi,t)  &= \tilde{T}_{31}^{EE}\,, \\
 F^{EF}(x,r,\xi,t)  & =  - 4 \tilde{T}_{31}^{EE} \,,\\
 F^{FF}(x,r,\xi,t)  & =4 \tilde{T}_{31}^{EE} + \frac{N_c}{16\pi^3} \frac{1}{\sigma_2^r}\bar {\cal C}_2(\sigma_2^{r}) \frac{\theta_{\bar\xi\xi}}{M^2\xi}\,,
\end{align}
\end{subequations}
with
\begin{align}
  \tilde{T}_{31}^{EE}&(x,r,\xi,t) =-  \frac{3 N_c}{16\pi^3} \frac{1}{\xi}  \int_0^1 d\alpha\,
\frac{\theta_{\alpha\xi}}{[\sigma_3^r]^2}\bar {\cal C}_3(\sigma_3^{r}) \,.
\end{align}

\subsection{$\mathbf{H_3^k}$}
\unskip
\begin{align}
H_3^k(x,& k_\perp^2,\xi,t)  = \bar P_{\rm T} \left[ E_\pi F_\pi \, F^{EF}  + F_\pi^2 \,F^{FF}\right] \,,
\label{EqGTMDH3l}
\end{align}
where
\begin{subequations}
\begin{align}
 F^{EF}(x,r,\xi,t)  & =  \tilde{T}_{41}^{EF}- \frac{N_c}{4\pi^3} \frac{1}{\sigma_2^r}\bar {\cal C}_2(\sigma_2^{r}) \frac{\theta_{\bar\xi\xi}}{M^2}\,,\\
 F^{FF}(x,r,\xi,t)  & =-2\tilde{T}_{41}^{EF} + \frac{N_c}{4\pi^3} \frac{1}{\sigma_2^r}\bar {\cal C}_2(\sigma_2^{r}) \frac{\theta_{\bar\xi\xi}}{M^2} \,,
\end{align}
\end{subequations}
with
\begin{align}
  \tilde{T}_{41}^{EF}& (x,r,\xi,t) = \nonumber \\
 & \frac{N_c}{8\pi^3}\frac{1}{M^2} \left[
  \frac{\theta_{\bar \xi 1}}{\sigma_1^{r,1}}\bar {\cal C}_2(\sigma_1^{r,1})
 - \frac{\theta_{\xi 1}}{\sigma_1^{r,-1}}\bar {\cal C}_2(\sigma_1^{r,-1}) \right].
\end{align}


\end{document}